\documentclass[a4paper,12pt]{article}
 \usepackage{graphicx}
 \DeclareGraphicsExtensions{.pdf,.png,.jpg}
 \pdfoutput=1
 \newcommand{\be}{\begin{equation}}
\newcommand{\ee}{\end{equation}}

\newcommand{\bea}{\begin{eqnarray}}
\newcommand{\eea}{\end{eqnarray}}

\newcommand{\p}{\partial}

\newcommand{\nn}{\nonumber \\}
\newcommand{\f}{\frac}

\newcommand{\ra}{\rightarrow}
\begin{document}

\thispagestyle{empty}
\begin{flushright}
{\bf arXiv:1312.0088}
\end{flushright}
\begin{center} \noindent \Large \bf Extremal Surfaces And Entanglement Entropy

\end{center}

\bigskip\bigskip\bigskip
\vskip 0.5cm
\begin{center}
{ \normalsize \bf   Shesansu Sekhar Pal }


\vskip 0.5 cm

Centre of Excellence in Theoretical and Mathematical Sciences, SOA University,  Bhubaneswar, 751030, India
\vskip 0.5 cm
\sf { shesansu${\frame{\shortstack{AT}}}$gmail.com }
\end{center}
\centerline{\bf \small Abstract}
We have obtained the  equation of the extremal  hypersurface   by considering the Jacobson-Myers functional  and computed the entanglement entropy.
In this context, we show that the higher derivative corrected extremal surfaces can not penetrate the horizon. 
Also, we have studied the entanglement temperature and entanglement entropy for low excited states for such higher derivative theories when the entangling region is of the strip type.

\newpage

\section{Introduction}

The study of the entanglement entropy in the AdS/CFT context \cite{Maldacena:1997re} has attracted a lot of attention  due to its potential application in condensed matter systems as well as in quantum information theories. In a seminal work, \cite{Ryu:2006bv},  Ryu and Takayanagi (RT) made a holographic  conjecture of the computation of the entanglement entropy. The proposal is given for a static spacetime with a co-dimension two hypersurface, whose area  is proposed to be related to the entanglement entropy. A proof of such a proposal is attempted in \cite{Fursaev:2006ih}, with further comments  in \cite{Headrick:2010zt}. In another study, a proof is  suggested in \cite{Casini:2011kv}, 
when the entangling region is of the sphere type. More recently, a suggestive  argument was put forward in  \cite{Lewkowycz:2013nqa} based on the previous works  in $1+1$ dimensional CFT of \cite{Hartman:2013mia} and \cite{Faulkner:2013yia} that explains the RT conjecture.

One of the universal result of  \cite{Ryu:2006bv} is that 
the entanglement entropy becomes divergent at UV, which goes as $\epsilon^{2-d}$, with $\epsilon$ as the UV regulator, for a $d$ dimensional CFT,  and becomes finite at IR. In fact, the approach adopted  in \cite{Ryu:2006bv} is a non-covariant way of doing the calculation, which  has been generalized in a covariant way  in \cite{Hubeny:2007xt} by Hubeny, Rangamani and Takayanagi 
(HRT) to derive, in particular, the equation of the extremal hypersurface. To emphasize, this is the generalization of the minimal hypersurface in a covariant way of RT prescription. 
However,  it is interesting to note that the equation of the hypersurface is also derived  in  \cite{Lewkowycz:2013nqa} starting from the bulk Einstein\rq{}s equation of motion, which is further studied in \cite{Bhattacharyya:2013jma,  Chen:2013qma} and \cite{Bhattacharyya:2013gra}. Some other  interesting studies  on entanglement entropy are  reported in \cite{Fischler:2012ca, Dong:2013goa} and \cite{Alishahiha:2013dca}.

In this paper, we ask several questions and find answers to it. First, how would the equation of the hypersurface look, in a covariant way, upon inclusion of the 
higher derivative terms to the entanglement entropy functional?
We answer this question, by considering the Jacobson-Myers functional (JM) \cite{Jacobson:1993xs} as the starting point for the holographic entanglement entropy functional and then derive the equation of the hypersurface by extremizing it with respect to the embedding fields\footnote{The embedding fields appear through the induced metric.}, $X^S$. It is important to note that such a from of the equation of the hypersurface does not depend on the shape or the size of the entangling region. 
As an example, with up to the Gauss-Bonnet term  in the  holographic entanglement entropy functional, the hypersurface reads as
 \bea\label{GB_eom}
&&   {\cal K}^S+\lambda_1\left(R{\cal K}^S-2R^{ab} {\cal K}^S_{ab}\right)+\Lambda\Bigg[ {\cal K}^S \left(R^2-4R_{a_1b_1}R^{a_1b_1}+ R_{a_1b_1c_1d_1} R^{a_1b_1c_1d_1}\right) -\nn&&4RR^{ab} {\cal K}^S_{ab}
+8R^{acbd}R_{cd} {\cal K}^S_{ab}-4R^{aecd}{R^b}_{ecd} {\cal K}^S_{ab}+8{R^a}_cR^{bc} {\cal K}^S_{ab}\Bigg]=0,
    \eea 
    where $ {\cal K}^S\equiv  g^{ab}{\cal K}^S_{ab}$, whose functional form is  given in eq(\ref{def_cal_K}). The couplings $\lambda_1$ and $\Lambda$ are undetermined constants and are dimensionfull objects. Let us recall, without turning on, $\Lambda$, the finite piece of the the $\lambda_1$  dependent part of the  entanglement entropy\footnote{Unless stated otherwise, throughout the paper, we compute the entanglement entropy for strip type and for $d\geq 3$.} is computed numerically  in \cite{deBoer:2011wk} for a $4$ dimensional CFT. We revisit such a  computation but  for a $d$ dimensional CFT, i.e., for bulk  AdS spacetime and the result to the linear order in the coupling $\lambda_1$ can be summarized as follows: (a) The divergent term coming from UV goes in the same way as in the absence of the higher derivative term to the entanglement entropy functional, i.e., like   $\epsilon^{2-d}$,   whereas  (b)  the finite term, the coefficient   of $r^{2-d}_{\star}$, where $r_{\star}$ is the turning point,  depends on the coupling $\lambda_1$ very non-linearly but we determine the functional form only to linear order. Explicitly,  it reads as
  \bea
  2 G_N&  S_{EE}&=\f{L^{d-2}R^{d-1}_0}{\epsilon^{d-2}}\left(\f{1-(d-1)(d-2)(\lambda_1/R^2_0)}{(d-2)}\right)-\nn&&L^{d-2}R^{d-1}_0\f{\sqrt{\pi}\Gamma\left(\f{d}{2(d-1)}\right)}{\Gamma\left(\f{1}{2(d-1)}\right)}r^{2-d}_{\star}\left(\f{1+(d-2)(d-3)(\lambda_1/R^2_0)}{d-2}\right)
  +{\cal O}(\lambda_1)^2\nn
  &=&\f{L^{d-2}R^{d-1}}{4\epsilon^{d-2}}\left(\f{4-(d+1)(d-1)(d-2)a\lambda_a}{(d-2)}\right)-\nn&&2^{d-2}L^{d-2}R^{d-1}\pi^{\f{d-1}{2}}\left(\f{\Gamma\left(\f{d}{2(d-1)}\right)}{\Gamma\left(\f{1}{2(d-1)}\right)}\right)^{d-1}\ell^{2-d}\left(\f{1+\f{3}{4}(d-1)(d-2)(d-3)a\lambda_a}{d-2}\right)
  \nn&&+{\cal O}(\lambda_a)^2,
  \eea

where $R_0$ and $R$ are the radii with and without the higher derivative term to the AdS and  $\lambda_1=a \lambda_a R^2$ with $a=\f{2}{(d-2)(d-3)}$. This value of $a$ follows upon comparing  our JM functional with eq(C.24) of \cite{Hung:2011xb} and\footnote{The value of $a$  makes sense for $d>3$, but in the limit of $d\ra 3$, the quantity $(d-3)a$ is finite.} identifying the coupling $\lambda_a=\lambda_{there}$.
After turning on both the couplings  $\lambda_1$ and $\Lambda$, we find that the entanglement entropy  for a $d=6$ dimensional CFT  
follows the same pattern. That is apart from the UV divergent term, $1/\epsilon^4$,  the finite term coming from IR  goes $1/r^4_{\star}$ times some constants. 
Explicitly, the entanglement entropy to linear order in the couplings  reads as
\bea
2 G_NS_{EE}&=&\f{L^4R^5_0}{4\epsilon^4}\left(1-\f{20\lambda_1}{R^2_0}+\f{120\Lambda}{R^4_0}\right)-\f{4L^4R^5_0}{11 \ell^4}\pi^{5/2}\left(11+\f{660\lambda_1}{R^2_0}-\f{600\Lambda}{R^4_0}\right)\left(\f{\Gamma\left(\f{3}{5} \right)}{\Gamma\left(\f{1}{10} \right)}\right)^5\nn
&=&\f{L^4R^5}{4\epsilon^4}\left(1-35a\lambda_a+300X_2\Lambda_a\right)-\nn&&\f{4L^4R^5}{11\ell^4}\pi^{5/2}\left(11+495 a\lambda_a-2460X_2\Lambda_a\right) \left(\f{\Gamma\left(\f{3}{5} \right)}{\Gamma\left(\f{1}{10} \right)}\right)^5.
 \eea

In fact, we need to set $a=1/6$ and $X_2=-1/24$. This follows upon comparing with eq(C.24) of \cite{Hung:2011xb} and identifying the couplings as
$\lambda_a=\lambda_{there}$ and $\Lambda_a=\mu_{there}$.

It simply follows from the expression of the entanglement entropy eq(\ref{ee_hsv})  or eq(\ref{area_exp_l_ads}) that there exists a differential equation relating the entanglement entropy and the size,  $\ell$, as
\be
\ell\f{\p}{\p\ell}\left(\ell \f{\p}{\p\ell}+(d-2) \right)S_{EE}=0,\quad {\rm with~fixed ~L~and~R }.
\ee

We ask the validity of such a differential equation upon inclusion of the higher derivative term to the holographic entanglement entropy functional? We check that such a form of the differential equation goes through to  linear order in the coupling $\lambda_1$ and $\Lambda$.
In fact, such a differential equation can be re-written as 
\be
\f{\p}{\p \ell}\left(\ell^{d-1}\f{\p}{\p\ell}S_{EE}\right)=0.
\ee

In the context of RG flow, in \cite{Myers:2012ed}, a candidate $c$-function has been  suggested, which shows the necessary monotonicity along the flow from UV to IR as long as the matter fields obeys the null energy condition\footnote{ This holds good only when the bulk action is of the  Einstein-Hilbert type but not for the Gauss-Bonnet type \cite{Myers:2012ed}.}. In fact, such a $c$-function\footnote{For $d=2$, the  $c$-function is defined in \cite{Casini:2003ix}.} is proportional to $\ell^{d-1}\f{\p}{\p\ell}S_{EE}$ for $d\geq 3$. The vanishing of the derivative of such a quantity with respect to the   size, $\ell$, suggests 
 that we are sitting at the fixed point.   However, in  \cite{Liu:2012eea} another quantity is defined and called as  \lq\lq{}renormalized entanglement entropy\rq\rq{} that gives the rate of flow\footnote{Such a quantity is defined strictly when the entangling region is of  the sphere type, which is UV finite. In our case,  with a slight abuse of notation, we define different quantities but use the same name and symbol as in  \cite{Liu:2012eea}.}. Inspired by this, we find   the simplest  UV finite  quantity is $\p_{\ell}S_{EE}$. Using which, we find there exists another quantity, $\ell^2\p^2_{\ell}S_{EE}-\ell \p_{\ell}S_{EE}$, which is negative but only when  $d\geq 5$ for non-zero $q$ in the perturbation to the geometry. 
 

In a recent study in \cite{Hubeny:2012ry} by Hubeny in  the context of the entanglement entropy  found that the spatial hypersurfaces does not penetrate the horizon. This motivate us to ask the following question:
Is the non-penetrating   of the horizon by the probe, spatial hypersurfaces, remain true even after the  inclusion of the higher derivative term to the area functional?  

The answer to this question  is  that upon inclusion of the higher derivative term of the type as in eq(\ref{assumption_EE}) to the action of the embedding field make the spatial hypersurface not to penetrate the horizon.



It is suggested  recently, by studying the low excited states  \cite{Bhattacharya:2012mi}, that the entanglement entropy obeys a law like that of the first law of thermodynamics, $T_{ent} \Delta S=\Delta E$.  The quantity $T_{ent}$ is dubbed as the \lq{}entanglement temperature\rq{}, which is proportional to   the inverse of the length, $\ell$, i.e., $T_{ent}=c~\ell^{-1}$. The proportionality constant, $c$,  is  a function of  $d$, the spacetime dimensionality of the field theory and depends on the nature of the entangling region \cite{Bhattacharya:2012mi}.  The question that we ask: Do we still expect to see  a similar first law like relation even after the inclusion of the higher derivative term to the holographic entanglement entropy functional? If yes, then how does the \lq{}entanglement temperature\rq{} go with the length, $\ell$? And how does  the proportionality constant  behave as a function of $d$? We show that there indeed exists a first law like relation even with the higher derivative term to the  holographic entanglement entropy functional and the relation between the \lq{}entanglement temperature\rq{} and the size is same  as mentioned above. And the proportionality constant is a function of $d$ and the coupling\footnote{Similar type of question is asked in \cite{Guo:2013aca} for a $4$ and $6$  dimensional CFT when the entangling region is of the strip type and sphere type, respectively. Our study is different in the following way.  We study the   $6$  dimensional CFT when the entangling region is of the strip type, which is not studied there. Also, we study the $d$ dimensional CFT by considering the two derivative area functional whose special case, $d=4$, is studied in \cite{Guo:2013aca}. }.

\bea
 c&=&\f{2(d^2-1)\left(\Gamma\left( \f{d}{2(d-1)}\right)\right)^2\Gamma\left( \f{d+1}{2(d-1)}\right)}{\sqrt{\pi}\left(\Gamma\left( \f{1}{2(d-1)}\right)\right)^2\Gamma\left( \f{1}{(d-1)}\right)}\left[1+4 \lambda_a \right],\quad {\rm for~any~d~with~\Lambda_a=0},\nn
 c&=& \f{70\Gamma\left(\f{7}{10}\right)\left( \Gamma\left(\f{3}{5}\right)\right)^2}{187\sqrt{\pi}\Gamma\left(\f{1}{5}\right)\left( \Gamma\left(\f{1}{10}\right)\right)^2} \left(187+748 \lambda_a+444\Lambda_a\right),\quad {\rm for~~d=6}.
\eea

The paper is organized as follows: In section 2 and Appendix A, we re-visit the computation of the area functional of the hyperscaling violating geometries and generates  both the existing \cite{Ogawa:2011bz, Huijse:2011ef, Dong:2012se}  and new form of the geometry that exhibits the log violation of the entanglement entropy. In section 3, we find the equation of the hypersurface using the technique of \cite{Hubeny:2007xt} but  with the higher derivative term to the area functional and explicitly compute the entanglement entropy for AdS spacetime. In section 4, we show following \cite{Hubeny:2012ry} the absence of the penetration of the horizon by the spatial hypersurfaces. 
  In section 5, we find the first law like of thermodynamics by considering fluctuation to the geometry and finally we conclude in section 6. Some of the expression of the solution  of the embedding field with the higher derivative term to the area functional has been relegated to Appendix B.

\section{A  differential equation}

If we look at the expression of the holographic entanglement entropy, which is proportional to the area, eq(\ref{area_exp_l_ads}) for AdS spacetime,  then it follows that 
\be\label{diff_ads_s_ee}
\ell\f{\p}{\p\ell}\left(\ell \f{\p}{\p\ell}+(d-2) \right)S_{EE}=0.
\ee
Note that in deriving such a differential  equation we have considered  a mild assumption that the UV cutoff, $\epsilon$, is independent of the size $\ell$, i.e., $\f{d\epsilon}{d\ell}=0$. In fact, this is true because the size, $\ell$ depends only on $r_{\star}$. The details of the computation of $S_{EE}$ are relegated to Appendix A.
This type of differential equation would suggest  some  form of the \lq{}RG flow\rq{} equation for the entanglement entropy.
 Strictly, such a form of the differential equation involving $S_{EE}$ and $\ell$ follows when the entangling region is of the strip type and holds true irrespective of whether $d$ is even or odd. 
In fact, a similar equation also follows for $r_{\star}$, as $\ell$ is linearly related to it.   It reads as
\be
r_{\star}\f{\p}{\p r_{\star}}\left(r_{\star} \f{\p}{\p r_{\star}}+(d-2) \right)S_{EE}=0.
\ee
It is expected that the limits of integration of the radial 
 direction should be independent of each other. Hence, it is justified to consider that $r_{\star}$ and $\epsilon$ are not dependent on each other.

The simple looking differential equation eq(\ref{diff_ads_s_ee})   obeyed by $S_{EE}$ for AdS spacetime gets changed when we change the background spacetime to Lifshitz type as follows
\be
\ell\f{\p}{\p\ell}\left(\ell \f{\p}{\p\ell}+(d-2-\gamma(d-1)) \right)S_{EE}=0.
\ee
This follows from eq(\ref{s_ee_lifshitz_area_functional}).
From which it follows that 
\be\label{c_function_lifshitz}
\f{\p}{\p\ell}\left(\ell^{(d-1)(1-\gamma)}\f{\p S_{EE}}{\p\ell} \right)=0.
\ee

 In \cite{Liu:2012eea}  a  quantity, ${\cal S}^{\Sigma}_d$, is defined with the shape of $\Sigma$  assumed to be a sphere. 
 This has been used to study the rate of flow of the renormalized entanglement entropy. In our case of a strip type entangling region, we shall assume the existence of  the following  quantity: $\ell \f{\p {\cal S}^{\Sigma}_d }{\p\ell}$,  
and   check  the consequence of imposition of $\ell \f{\p {\cal S}^{\Sigma}_d }{\p\ell} < 0$ on  $S_{EE}$ for $d=3$ and $4$.
  To make it clear, we have taken the following forms, ${\cal S}^{\Sigma}_3=\ell \p_{\ell} S_{EE}-S_{EE}$ and  ${\cal S}^{\Sigma}_4=\f{\ell}{2}\p_{\ell}(\ell\p_{\ell}S_{EE}-2S_{EE})$. Note that ${\cal S}^{\Sigma}_3$ is not UV finite unlike ${\cal S}^{\Sigma}_4$.

On imposition of $\ell \f{\p {\cal S}^{\Sigma}_d }{\p\ell} < 0$ for $d=3,~4$, we find
\be
\ell \p_l S_{EE}> 0~{\rm for~d=3},\quad {\rm and}~\ell \p_l S_{EE}< 0~{\rm for~d=4}.
\ee

In getting such a result, we have used eq(\ref{diff_ads_s_ee}). Upon evaluating $\ell \p_l S_{EE}$ explicitly, using eq(\ref{area_exp_l_ads}),  we find it to be positive for any $d$. For $d=3$, this result is in some sense consistent with \cite{Liu:2012eea} because the authors  did not find any example that violates the monotonicity of ${\cal S}_3$ unlike ${\cal S}_4$.

In what follows, we shall define two more quantities, which are UV finite, $\ell\p_{\ell}\left(\ell\p_{\ell}S_{EE}\right)$ and $\ell^3\p_{\ell}\left(\ell^{-1}\p_{\ell}S_{EE}\right)$ and find whether they can be used to study the rate of flow, i.e., are negative?
Given the explicit result of the entanglement entropy for AdS spaetime in eq(\ref{area_exp_l_ads}), we find the following equations for any $d\geq 3$
\be\label{plausible_flow_any_d}
\ell\p_{\ell}\left(\ell\p_{\ell}S_{EE} \right)=\ell^2\p^2_{\ell}S_{EE}+\ell \p_{\ell}S_{EE}<0,\quad \ell^3\p_{\ell}\left(\ell^{-1}\p_{\ell}S_{EE} \right)=\ell^2\p^2_{\ell}S_{EE}-\ell \p_{\ell}S_{EE}<0.
\ee

In fact there exists the following relation
\be
\ell \p_{\ell}{\cal S}^{\Sigma}_3=\f{\ell\p_{\ell}\left(\ell\p_{\ell}S_{EE} \right)+\ell^3\p_{\ell}\left(\ell^{-1}\p_{\ell}S_{EE} \right)}{2}.
\ee
Hence, it is not surprising to see that $\ell \f{\p {\cal S}^{\Sigma}_3 }{\p\ell} < 0$. Hence, it is highly plausible to consider  these two quantities, which  may give information on the rate of flow.
This is further investigated by doing fluctuations to the geometry  in subsection 6.3 and there arises interesting restrictions on $d$.

\subsection{Log term to $S_{EE}$ }

In this section, we shall try to find  answer to the following question: Can we generate $log$ term in the entanglement entropy? 
 Let us assume the following form of the metric in $3+1$ dimensional spacetime
\bea\label{generic_spacetime}
ds^2_{3+1}&=&-g_{tt}(r)dt^2+g_{xx}(r)dx^2+g_{yy}(r)dy^2+g_{rr}(r)dr^2+2 g_{xy}(r) dxdy,
\eea
we assume that the boundary is at $r=0$.
Following the proposal of \cite{Ryu:2006bv}, the geometry of the  co-dimension two   hypersurface becomes
\be
ds^2_2=\left(g_{xx}+g_{rr}~r\rq{}^2(x)\right)dx^2+g_{yy}dy^2+2 g_{xy}dxdy
\ee
 
So the area of the hypersurface becomes
\be
{\cal A}=2\int dy dr \sqrt{(g_{xx}g_{yy}-g^2_{xy})~x\rq{}^2(r)+g_{rr}g_{yy}},
\ee
where we have inverted the function $r(x)$ and written as $x(r)$. Now the hypersurface $x(r)$ that extremizes the area ${\cal A}$ is
\be\label{velocity}
\f{dx}{dr}=\f{\sqrt{(g_{xx}(r_{\star})g_{yy}(r_{\star})-g^2_{xy}(r_{\star}))}\sqrt{ g_{yy}(r)g_{rr}(r)}}{\sqrt{[g_{xx}(r)g_{yy}(r)-g^2_{xy}(r)][g_{xx}(r)g_{yy}(r)-g^2_{xy}(r)-g_{xx}(r_{\star})g_{yy}(r_{\star})+g^2_{xy}(r_{\star})]}},
\ee
 where the turning point $r_{\star}$ is determined when the quantity $x\rq{}(r_{\star})$ diverges. We consider the 
 entangling region to be like a strip, $0 \leq x \leq \ell$ and $-L/2\leq y\leq L/2$ and the size 
 \be\label{length_x}
 \ell/2=\int^{r_{\star}}_0dr \f{\sqrt{(g_{xx}(r_{\star})g_{yy}(r_{\star})-g^2_{xy}(r_{\star}))}\sqrt{ g_{yy}(r)g_{rr}(r)}}{\sqrt{[g_{xx}(r)g_{yy}(r)-g^2_{xy}(r)][g_{xx}(r)g_{yy}(r)-g^2_{xy}(r)-g_{xx}(r_{\star})g_{yy}(r_{\star})+g^2_{xy}(r_{\star})]}} .
 \ee
In such  a case, the extrmized area turns out to be
\be\label{area_EE}
{\cal A}=2L\int^{r_{\star}}_{\epsilon} dr \f{\sqrt{g_{rr}(r)g_{yy}(r)}}{\sqrt{1-\f{g_{xx}(r_{\star})g_{yy}(r_{\star})-g^2_{xy}(r_{\star})}{g_{xx}(r)g_{yy}(r)-g^2_{xy}(r)}}},
  \ee  
where $\epsilon$ is the UV-cutoff. From  now on wards, we shall be working in the diagonal form of the bulk metric i.e.,  $g_{xy}=0$, for simplicity.  From this expression of the area, we can ask: under what condition do we see a log term?
The condition to see such a log term, using eq(\ref{integrals})  are
\be
g_{rr}g_{yy}=\f{1}{r^{2(2k+1)}},\quad g_{xx}g_{yy}=\f{1}{r^2},\quad {\rm for}\quad k=0,1,2,\cdots.
\ee

 In which case the area becomes
\be\label{gen_exp_area}
{\cal A}(k)=2L R^2\int^{r_{\star}}_{\epsilon} \f{dr}{r^{2k+1}}\f{1}{\sqrt{1-(r/r_{\star})^{2}}},\quad {\rm for}\quad k=0,1,2,\cdots.
\ee
We have put an index $k$ in the area to label it.  
In order to fix the metric components, let us assume that 
\be
g_{yy}=r^{-2w},\quad g_{xx}=r^{2(w-1)},\quad g_{rr}=r^{2(w-2k-1)},
\ee
in which case the $3+1$ dimensional geometry reads as
\be
ds^2_{3+1}(k,w)=R^2\left[-g_{tt}(r)dt^2+\f{dx^2}{r^{2(1-w)}}+\f{dy^2}{r^{2w}}+\f{dr^2}{r^{2(2k+1-w)}}\right],
\ee
where we have used  the indices $k$ and $w$ to label the geometry. Moreover, 
the entanglement entropy does not fixes the time-time component of the metric tensor. The quantity 
\be\label{gen_exp_ell}
\ell/2=\f{1}{r_{\star}}\int^{r_{\star}}_0 \f{dr}{r^{2k-1}}\f{1}{\sqrt{1-(r/r_{\star})^{2}}}.
\ee

Note, there   exists  UV divergence to $\ell$ for $k \geq 1$. Once again to regulate it, we have put an UV cutoff. In which case, 
\be
\ell/2=\f{1}{r_{\star}}\int^{r_{\star}}_{\epsilon} \f{dr}{r^{2k-1}}\f{1}{\sqrt{1-(r/r_{\star})^{2}}},\quad {\rm for}\quad k\geq 1 .
\ee

Explicitly, the area for $k=0$ and $k=1$ are
\bea
{\cal A}(k=0)&=&2LR^2 ~Log\left( \f{2r_{\star}}{\epsilon}\right),\quad  \ell/2=r_{\star},\nn
{\cal A}(k=1)&=&2LR^2 ~\left[\f{1}{2r^2_{\star}} Log \left( \f{2r_{\star}}{\epsilon}\right)-\f{1}{4r^2_{\star}}+\f{1}{2\epsilon^2} \right],\quad \ell/2=\f{1}{r_{\star}}~Log ~\left(\f{2r_{\star}}{\epsilon}\right) 
\eea
Note for $k=0$ case, the entanglement entropy obeys eq(\ref{c_function_lifshitz}) for $\gamma=1/2$.

It is interesting to note that the neither the area eq(\ref{gen_exp_area}) nor the length, $\ell$, eq(\ref{gen_exp_ell}) depends on  $w$. It means there can be more than one co-dimension two geometry which has the same  area and the length $\ell$.
The explicit from of the bulk geometry for $k=0,~1$ are
\bea
ds^2_{3+1}(k=0,w)&=&R^2\left[-g_{tt}(r)dt^2+\f{dx^2}{r^{2(1-w)}}+\f{dy^2}{r^{2w}}+\f{dr^2}{r^{2(1-w)}}\right],\nn
ds^2_{3+1}(k=1,w)&=&R^2\left[-g_{tt}(r)dt^2+\f{dx^2}{r^{2(1-w)}}+\f{dy^2}{r^{2w}}+\f{dr^2}{r^{2(5-w)}}\right]
\eea

For $k=1$ case, it is very difficult to find the explicit dependence of  the area in terms of $\ell$ and the UV cutoff $\epsilon$. Even though the area depends logarithmically on $r_{\star}$, but it is not clear it will do so in terms of $\ell$. Also,  equation of the type,  eq(\ref{c_function_lifshitz}), is difficult to satisfy for $\gamma=3/2$, where $\gamma$ is defined below in eq(\ref{scale_with_w_1/2}).

Looking at the geometries for $k=0$ and $1$, we see the presence of rotational symmetry only for $w=1/2$. Hence, let us find out the scaling behavior of such cases.

\paragraph{Case 1:~ $w=1/2$.}

 For this choice of $w$,  there exists a rotational symmetry, i.e.,  $
g_{xx}=g_{yy}$. 
The choice $k=0$ corresponds to that studied in the previous section and also found in \cite{Huijse:2011ef}. The other choices of $k \geq 1$ are new. 

As an example, let us looks at the explicit  geometry for $k=1$. In which case, we get
\be
g_{xx}=g_{yy}=\f{R^2}{r},\quad  g_{rr}=\f{R^2}{r^5}.
\ee


It means, we can write down the geometry as
\bea
ds^2_4&=&R^2\left[-g_{tt}(r)dt^2+\f{(dx^2+dy^2)}{r}+\f{dr^2}{r^{4k+1}} \right]\nn
&=& R^2 r \left[-{\tilde g}_{tt}(r)dt^2+\f{(dx^2+dy^2)}{r^2}+\f{dr^2}{r^{2(2k+1)}} \right],
\eea
which falls under the category of the hyperscaling violating geometry provided $t$ has a nice scaling behavior. It is easy to see that for $k=0$  the geometry is  same as written in eq(\ref{geometry_hsv}) for $d=3$ and $\gamma=1/2$. If we assume that  $g_{tt}=\rho^{\f{1-4k+2z(1-2k)}{2k-1}}$, then for generic $k$ we can have   the following scaling behavior 
\be\label{scale_with_w_1/2}
\rho \ra \f{\rho}{\lambda},\quad t\ra \lambda^{-z}~ t,\quad x^i\ra \lambda ~x^i,\quad ds\ra \lambda^{\f{4k-1}{2(2k-1)}}~ds\equiv \lambda^{\gamma}~ds,
\ee
where $\rho=r^{2k-1}$.

\paragraph{Case 2:}

This corresponds to those solutions which does not respect the rotational symmetry i.e., solutions other than $w=1/2$. Let us take different choices  of $w$. In which case, the $3+1$ dimensional solution becomes
\bea\label{non_rotational_log_term_s_ee}
ds^2_{3+1}&=&R^2\left[-g_{tt}(r)dt^2+\f{dx^2}{r^2}+dy^2+\f{dr^2}{r^{2(2k+1)}}\right],\quad {\rm for}\quad  w=0\nn
&=&R^2\left[-g_{tt}(r)dt^2+\f{dy^2}{r^2}+dx^2+\f{dr^2}{r^{4k}}\right],\quad {\rm for}\quad  w=1.
 \eea 

For $g_{tt}=r^{-2}$ with $k=0$ and $w=0$,  it corresponds to $AdS_3\times R^1$ or $AdS_3\times S^1$ for non-compact and compact, $y$, respectively. As we saw earlier, the entanglement entropy does not depend on $w$ and for $k=0$ case it goes as $S_{EE}\sim Log(\ell/\epsilon)$. It means for $k=0$ and $w=0$, we can have a log violation to the entanglement entropy as well.

\paragraph {Subsummary:} We obtain  the presence of the logarithmic term  in the entanglement entropy in $3+1$ dimensional bulk system with the rotational symmetry along the spatial directions.  In fact the only geometry that is found here, corresponding to having $k=0$, and is same as that found  in  \cite{Huijse:2011ef}, which is of the Lifshitz type. So, the only rotationally invariant solution whose entanglement entropy goes logarithmically with the size, $\ell$, corresponds to the $k=0$ case.

In the absence of the  rotational symmetry, we have obtained several geometries as written in eq(\ref{non_rotational_log_term_s_ee}). In particular, the direct product of geometries like  $AdS_3\times R^1$ or $AdS_3\times S^1$ shows the presence of log term in $S_{EE}$.

\subsection{Re-visit of  $S_{EE}$ as studied in  \cite{deBoer:2011wk}}

In \cite{deBoer:2011wk}, the authors considered a  geometry that does not have the full rotational symmetry SO(d-1) in a $d+1$ dimensional bulk system,  while studying the entanglement entropy. 
 We are going to re-investigate this calculation  but without the
higher derivative correction.

Let us substitute in eq(\ref{area_EE}) the following form of the metric components as considered in  \cite{deBoer:2011wk}
\be
g_{tt}=g_{xx}=g_{rr}=1/r^2,\quad g_{yy}=1/r^{2w},\quad g_{xy}=0.
\ee

Now the hypersurface is described by $x(r)$, whose explicit  form can be read out from eq(\ref{velocity}). On computing the area of the hypersurface of a strip type entangling region 
\bea
\f{{\cal A}}{2}&=&L \int^{r_{\star}}_{\epsilon} dr \f{1}{r^{w+1}\sqrt{1-\left(\f{r}{r_{\star}} \right)^{2(w+1)}}}=-\left(\f{L}{wr^w}{}_2F_1\left[\f{1}{2},-\f{w}{2(w+1)},\f{w+2}{2(w+1)},\left(\f{r}{r_{\star}}\right)^{2(w+1)} \right]\right)^{r_{\star}}_{\epsilon}\nn
&=& \f{L}{w\epsilon^w}-\f{L\sqrt{\pi}}{(2w+1)r^w_{\star}}\f{\Gamma\left(-\f{w}{2(w+1)} \right)}{\Gamma\left(-\f{2w+1}{2(w+1)} \right)}.
\eea
 and the size 
 \be
 \ell/2=r_{\star}\sqrt{\pi} \f{\Gamma\left(\f{2+w}{2(1+w)}\right)}{\Gamma\left(\f{1}{2(1+w)}\right)}.
 \ee
Note that the $3+1$ dimensional  geometry as  written above is a solution at IR not at UV. Hence, it is expected that such a solution will not  show the desired result,  $\epsilon^{-1}$, at UV. In such a case, the entanglement entropy obeys the following differential equation
\be
\ell\f{\p}{\p\ell}\left(\ell \f{\p}{\p\ell}+w \right)S_{EE}=0.
\ee
 
\section{The equation of the hypersurface}
 
 In this section, we shall derive the covariant equation of the extremal hypersurface with higher derivative effects. It means, we are including  the effects of the finite \lq{}t Hooft coupling. The  equation motion of the embedding fields, $X^M(\sigma^a)$ essentially gives the form of the hypersurface. The induced metric is given by $g_{ab}=\f{\p X^M}{\p\sigma^a} \f{\p X^N}{\p\sigma^b} G_{MN} $, where $G_{MN}$ denotes the $d+1$  dimensional geometry of the bulk spacetime, $\sigma^a$ are the coordinates on the codimension-2 hypersurface. Let us assume that the entanglement entropy functional is
 \be\label{assumption_EE}
4 G_N S_{EE}=\int d^{d-1}\sigma~\sqrt{det( g_{ab})}\left[1+\lambda_1 R(g)+\lambda_2 R^2(g)+\lambda_3 R_{ab}(g)R^{ab}(g)+\lambda_4 R_{abcd}(g)R^{abcd}(g) \right]
 \ee
 where we have included higher derivative correction with $\lambda_i$\rq{}s as the coefficients\footnote{In \cite{Ogawa:2011fw}, another kind of higher derivative term in the  entanglement entropy functional is studied.}. 
 
{\it A priori} there is no good reason to believe  the inclusion of  higher derivative terms in this particular way. Even though this  is purely a guess but we hope,  it can be thought of as follows. For the Einstein-Hilbert action, it is suggested in \cite{Ryu:2006bv} to consider only the first term in eq(\ref{assumption_EE}), i.e., setting all the $\lambda_i$\rq{}s to zero. Upon inclusion of the   Gauss-Bonnet term to the (bulk)  Einstein-Hilbert action, it is suggested in  \cite{deBoer:2011wk}  that the entanglement entropy should be  given by eq(\ref{assumption_EE}) for which $\lambda_2=0=\lambda_3=\lambda_4$. Hence, it follows from these examples that for each power of the \lq{}Ricci scalar\rq{} in the bulk theory\footnote{Here we mean by \lq{}Ricci scalar\rq{} are the terms with all possible combination of Riemann tensor, Ricci tensor as well as Ricci scalar that is diffeo  invariant.} one should include a \lq{}Ricci scalar\rq{} with one less power in the entanglement entropy. Note that the \lq{}Ricci scalar\rq{} in the entanglement entropy should be constructed out of the induced metric $g_{ab}$. Formally, we  can write it  as 
  \be
 \int \sqrt{G} \left[{\cal R}(G)+\f{d(d-1)}{R^2}\right]\longrightarrow \int \sqrt{g},~  \int \sqrt{G}~\left[{\cal R}(G)+ \lambda{\cal GB}(G)\right] \longrightarrow \int \sqrt{g}\left[1+2\lambda R(g)\right],
  \ee
  where the left hand side is the bulk action and right hand side is the entanglement entropy functional and ${\cal GB}(G)$ denotes the Gauss-Bonnet term made from the bulk metric $G_{MN}$. Similarly, if we go for one more higher power of the scalar curvature then it is highly plaussible to consider the terms as written in   eq(\ref{assumption_EE}) with arbitrary coefficients. In \cite{Hung:2011xb}, the authors have considered the Jacobson-Myers form of the entropy functional \cite{Jacobson:1993xs} and studied the entanglement entropy when the entangling region is of the sphere and cylinder type. In our study, we do it for the strip type. Before moving onto the calculation of the extremal surface, recently in \cite{Dong:2013goa}, the author has given a derivation of the Jacobson-Myers entropy functional. 

On varying the entanglement entropy functional,  eq(\ref{assumption_EE}),  with respect to the embedding field, $X^S$, gives 
\be\label{def_cal_K}
X^{ab}{\cal K}^S_{ab}+\p_b X^S \nabla_a X^{ab}=0,\quad {\rm where}\quad {\cal K}^S_{ab}\equiv \p_a\p_b X^S-\gamma^c_{ab} \p_c X^S+\p_a X^M\p_b X^N \Gamma^S_{MN}
 \ee
where the  $\gamma^c_{ab}$ and $\Gamma^K_{MN}$ are connections defined with respect to $g_{ab}$ and $G_{MN}$, respectively. In fact, 
${\cal K}^S_{ab}$ obeys an identity ${\cal K}^M_{ab} \p_c X^N G_{MN}=0$. 
The quantity 
\bea
X^{ab}&=&\f{1}{2}g^{ab}+\lambda_1\left(\f{1}{2}g^{ab} R-R^{ab} \right)+\lambda_2\left(\f{1}{2}g^{ab}R^2-2RR^{ab}+\nabla^a\nabla^b R+\nabla^b\nabla^a R-2g^{ab}\nabla^2 R \right)\nn
&+&\lambda_3\left(\f{1}{2}g^{ab}R_{cd}R^{cd} +\f{1}{2}\nabla^a\nabla^b R+\f{1}{2}\nabla^b\nabla^a R-\f{1}{2}g^{ab}\nabla^2 R-2R^{acbd} R_{cd}-\nabla^2 R^{ab}\right)\nn&+&
\lambda_4\Bigg(\f{1}{2}g^{ab} R_{a_1b_1c_1d_1} R^{a_1b_1c_1d_1}-2 R^{aecd}{R^b}_{ecd}-4\nabla^2 R^{ab} +\nabla^a\nabla^b R+\nabla^b\nabla^a R-4R^{acbd}R_{cd}\nn&+&4R^{a}_cR^{bc}\Bigg).
\eea
   
  For this form of $X^{ab}$, one can easily show that, it obeys an identity: $\nabla_a X^{ab}=0$, where the covariant derivative is defined with respect to $g_{ab}$. In which case, the equation of motion of the extremal hypersurface becomes 
  \be\label{generic_hyper_surface_EE}
  X^{ab}{\cal K}^S_{ab}=0.
  \ee
  Note, the equation of the hypersurface is independent of the shape and size of the  entangling region.
  Now, we shall write down the form of $X^{ab}$ in two different cases.
  
  \paragraph{ Gauss-Bonnet combination:}

Let us consider a very  specific combination where $\lambda_2=\Lambda,~    \lambda_3=-4\Lambda,~\lambda_4=\Lambda$, then $X^{ab}$ takes the following form
\bea
X^{ab}&=&\f{1}{2}g^{ab}+\lambda_1\left(\f{1}{2}g^{ab} R-R^{ab} \right)+\Lambda\Bigg[\f{1}{2}g^{ab} \left(R^2-4R_{a_1b_1}R^{a_1b_1}+ R_{a_1b_1c_1d_1} R^{a_1b_1c_1d_1}\right) \nn&&-2RR^{ab}+4R^{acbd}R_{cd}-2R^{aecd}{R^b}_{ecd}+4R^{a}_cR^{bc}\Bigg].
\eea
  
   In which case, the equation of motion of $X^S$ can be re-written as
   \bea\label{GB_eom}
&&   {\cal K}^S+\lambda_1\left(R{\cal K}^S-2R^{ab} {\cal K}^S_{ab}\right)+\Lambda\Bigg[ {\cal K}^S \left(R^2-4R_{a_1b_1}R^{a_1b_1}+ R_{a_1b_1c_1d_1} R^{a_1b_1c_1d_1}\right) -\nn&&4RR^{ab} {\cal K}^S_{ab}
+8R^{acbd}R_{cd} {\cal K}^S_{ab}-4R^{aecd}{R^b}_{ecd} {\cal K}^S_{ab}+8{R^a}_cR^{bc} {\cal K}^S_{ab}\Bigg]=0,
    \eea 
    where $ {\cal K}^S\equiv  g^{ab}{\cal K}^S_{ab}$.
    
     \paragraph{ Weyl-square combination:}
    
      In this case, the $\lambda_i$\rq{}s take the following values: $\lambda_2=\f{2\Lambda}{(d-2)(d-3)},~\lambda_3=-\f{4\Lambda}{d-3}$ and $\lambda_4=\Lambda$. In which case the $X^{ab}$ takes the following form 
      
      \bea
      X^{ab}&=&\f{1}{2}g^{ab}+\lambda_1\left(\f{1}{2}g^{ab} R-R^{ab} \right)+\Lambda\Bigg[\f{1}{2}g^{ab}\bigg(\f{2}{(d-2)(d-3)}R^2-\f{4}{d-3}R_{a_1b_1}R^{a_1b_1}+\nn &&
       R_{a_1b_1c_1d_1} R^{a_1b_1c_1d_1}\bigg) -\f{4}{(d-2)(d-3)}RR^{ab}+\f{d-4}{d-2}\nabla^a\nabla^bR+\f{d-4}{d-2}\nabla^b\nabla^a R+\nn&&
       \f{2(d-4)}{(d-2)(d-3)}g^{ab}\nabla^2 R+4\f{(d-5)}{d-3}R^{acbd}R_{cd}-4\f{(d-4)}{d-3}\nabla^2 R^{ab}-2R^{aecd}{R^b}_{ecd}+\nn&&4R^{a}_cR^{bc}\Bigg]
      \eea
   and the equation of motion of $X^S$ becomes
     \bea
&&   {\cal K}^S+\lambda_1\left(R{\cal K}^S-2R^{ab} {\cal K}^S_{ab}\right)+\Lambda\Bigg[ {\cal K}^S  \bigg(\f{2}{(d-2)(d-3)}R^2-\f{4}{d-3}R_{a_1b_1}R^{a_1b_1}+\nn &&
       R_{a_1b_1c_1d_1} R^{a_1b_1c_1d_1}\bigg) -\f{8}{(d-2)(d-3)}RR^{ab}{\cal K}^S_{ab}+2\f{(d-4)}{d-2}{\cal K}^S_{ab}\left(\nabla^a\nabla^bR+\nabla^b\nabla^a R\right)+\nn&&
       \f{4(d-4)}{(d-2)(d-3)}{\cal K}^S\nabla^2 R+8\f{(d-5)}{d-3}R^{acbd}R_{cd}{\cal K}^S_{ab}-8\f{(d-4)}{d-3}{\cal K}^S_{ab}\nabla^2 R^{ab}-\nn&&4R^{aecd}{R^b}_{ecd}{\cal K}^S_{ab}+8{\cal K}^S_{ab}R^{a}_cR^{bc}\Bigg]=0.
    \eea 
    
    We have checked that  the  Weyl-squared term does not contribute to  the entanglement entropy till $d=8$. 
    
   \subsection{The precise form of the hypersurface: An example for strip} 
   
 Let us compute the   hypersurface eq(\ref{generic_hyper_surface_EE}), for the following form of the  solution in the bulk 
   \be
   ds^2_{d+1}=G_{MN}dx^Mdx^N=-g_{tt}(r)dt^2+g_{xx}(r)\left( dx^2_1+\cdots+dx^2_{d-1}\right)+g_{rr}(r)dr^2
   \ee
   which gives rise to the following induced metric with the embeddings as
   \bea\label{embeddings_strip}
&& X^t=0,\quad X^a=x^a=\sigma^a,\quad X^r=r(x_1);\nn && 
ds^2_{d-1}=g_{ab}d\sigma^ad\sigma^b= \left[g_{xx}(r)+g_{rr}(r)r\rq{}^2\right]dx^2_1+g_{xx}(r)\left( dx^2_2+\cdots+dx^2_{d-1}\right),
   \eea
   where $r\rq{}=\f{dr}{dx_1}$. The strip is extended along $0 \leq x_1 \leq \ell,~ -L/2\leq (x_2,\cdots,x_{d-1})\leq L/2$.
  The explicit computation of the components of ${\cal K}^S_{ab}$ gives
   \bea
   {\cal K}^{x_1}_{x_1x_1}&=&\f{r\rq{}g_{xx}\p_r g_{xx}+2r\rq{}^3g_{rr}\p_r g_{xx}-r\rq{}^3g_{xx}\p_r g_{rr}-2r\rq{}r\rq{}\rq{}g_{xx}g_{rr}}{2g_{xx}(g_{xx}+r\rq{}^2 g_{rr})},\nn
   {\cal K}^{x_1}_{x_ix_j}&=& \f{r\rq{}\p_r g_{xx}}{2(g_{xx}+r\rq{}^2 g_{rr})}\delta_{x_ix_j},\quad  {\cal K}^{r}_{x_ix_j}=-\f{g_{xx}\p_r g_{xx}}{2g_{rr}(g_{xx}+r\rq{}^2 g_{rr})}\delta_{x_ix_j},~ (i,j=2,\cdots,d-1)\nn
  {\cal K}^r_{x_1x_1}&=&\f{2g_{rr}g_{xx}r\rq{}\rq{}-2r\rq{}^2g_{rr}\p_rg_{xx}+r\rq{}^2g_{xx}\p_r g_{rr}-g_{xx}\p_r g_{xx}}{2g_{rr}(g_{xx}+r\rq{}^2 g_{rr})}  
    \eea 
and the rest of the components are zero.   For simplicity, let us set the coefficients $\lambda_i$\rq{}s to zero in which case,  the extremal    
   hypersurface, ${\cal K}^S=0$, gives
   \be
 2r\rq{}\rq{} g_{xx}g_{rr}-dr\rq{}^2 g_{rr}\p_r g_{xx}+r\rq{}^2 g_{xx}\p_r g_{rr}-(d-1) g_{xx}\p_r g_{xx}=0.
   \ee

       Upon using the identity, $r\rq{}^3 \f{d^2x_1}{dr^2}=-r\rq{}\rq{}$, we can re-write the equation of the  hypersurface as 
   \be
 2x_1\rq{}\rq{} g_{xx}g_{rr}+d g_{rr}\p_r g_{xx}x_1\rq{}- g_{xx}\p_r g_{rr}x_1\rq{}+(d-1) g_{xx}\p_r g_{xx}x\rq{}^3_1=0\Longrightarrow \f{d}{dr}\left(\f{g^{\f{d}{2}}_{xx}x_1\rq{}}{\sqrt{g_{rr}+g_{xx}x_1\rq{}^2}} \right)=0,
   \ee
   where $x\rq{}_1=\f{dx_1}{dr}$. Essentially, we have re-written a second differential equation as a first order differential equation.   Now, we can solve the equation of motion and 
   \be\label{sol_strip}
   \f{dx_1}{dr}=\f{g^{\f{d-1}{2}}_{xx}(r_{\star})\sqrt{g_{rr}(r)}}{\sqrt{g^{d}_{xx}(r)-g_{xx}(r)g^{d-1}_{xx}(r_{\star})}}.
   \ee
  In order to determine the constant of integration, we have used  the following boundary condition, $x_1\rq{}(r_{\star})\ra\infty$.
  To get a feel of the solution,  let us consider a spacetime that exhibits the scale violating behavior along with the trivial and non-trivial scaling of the spatial direction \cite{Pal:2012zn}, namely, 
   \be\label{generic_hsv}
   ds^2_{d+1}=r^{2\gamma}\left[-\f{dt^2}{r^{2z}}+\f{dx^2_1+\cdots+dx^2_{d-1}}{r^{2\delta}}+\f{dr^2}{r^2}\right],\quad~{\rm where}~ \delta=0~ {\rm or}~ 1~{\rm with}~\gamma \leq 0.
   \ee
   In which case the differential equation can be exactly solved 
   \bea
  \pm x_1(r)&=&c_1+\f{r^{d\delta-\gamma(d-1)}}{r^{(d-1)(\delta-\gamma)}_{\star}}\f{1}{[d\delta-\gamma(d-1)]}\times\nn&&{}_2F_1\left[ \f{1}{2},\f{d\delta-\gamma(d-1)}{2(d-1)(\delta-\gamma)},\f{\delta(3d-2)-\gamma(d-1)}{2(d-1)(\delta-\gamma)},\left(\f{r}{r_{\star}}\right)^{2(d-1)(\delta-\gamma)}\right],
   \eea
   where $c_1$ is a constant of integration and ${}_2F_1[a,b,c,x]$ is the hypergeometric function. The precise form of $c_1$ is determined by imposing the boundary condition that $x_1(r=r_{\star})=0$ \cite{Hubeny:2012ry}, which gives
   \be
   c_1=-r^{\delta}_{\star}\f{\sqrt{\pi}}{\delta}\f{\Gamma\left(\f{d\delta-\gamma(d-1)}{2(d-1)(\delta-\gamma)}\right)}{\Gamma\left(\f{\delta}{2(d-1)(\delta-\gamma)}\right)}.
   \ee
  
   We can relate   $\ell$  with the constant of integration, $c_1$, as $ \ell/2=-c_1$. 
   It is easy to see that in the $\delta=1,~\gamma=0$  limit, it re-produces the result of \cite{Ryu:2006bv}. However, in the  $\delta=0$ and $\gamma=0$ limit, the solution becomes
   \be
   \pm x_1(r)=c_1-\f{1}{\gamma(d-1)} Sin^{-1}\left(\f{r_{\star}}{r} \right)^{\gamma(d-1)},\quad c_1=\f{\pi}{2\gamma(d-1)}.
   \ee

   The entanglement entropy for a generic diagonal and rotationally invariant metric  with the entangling region as a strip, $0 \leq x_1 \leq \ell,~-L/2\leq (x_2,\cdots,x_{d-1})\leq L/2, $ takes the following form \cite{Pal:2012zn}
   \be\label{s_ee_leading_order}
2 G_N   S_{EE}=L^{d-2}\int^{r_{\star}}_{\epsilon} dr \f{\sqrt{g^{d-2}_{xx}(r)g_{rr}(r)}}{\sqrt{1-\left(\f{g_{xx}(r_{\star})}{g_{xx}(r)}\right)^{d-1}}}=L^{d-2}\int^{r_{\star}}_{\epsilon} dr \f{r^{(\gamma-\delta)(d-2)+\gamma-1}}{\sqrt{1-\left(\f{r_{\star}}{r}\right)^{2(\gamma-\delta)(d-1)}}},
   \ee
   where in the second equality we have substituted the geometry as written in eq(\ref{generic_hsv}). Now, we shall give results to this integral in two different cases i.e., $\delta=0,~1$ and $\gamma\neq \delta$.
   
\paragraph{  For  $ \delta=0,~\gamma\neq 0~{\rm case}:\quad$}  In this case the entanglement entropy gives the following result
   \be 
2 G_N   S_{EE}(\delta=0)=-L^{d-2}\f{\epsilon^{\gamma(d-1)}}{\gamma(d-1)}~\sqrt{1-\left(\f{r_{\star}}{\epsilon}\right)^{2\gamma(d-1)}}.
   \ee
     It looks from this expression as if the area is a complex quantity but it is not because $\gamma $ is negative.  The entanglement entropy is completely divergent and the divergence goes as $\epsilon^{-|\gamma|(d-1)}$.
     
\paragraph{  For   $\delta=1,~\gamma\neq 0~{\rm case}:\quad$} In this case the entanglement entropy  gives the following result
    \be
 2 G_N     S_{EE}(\delta=1)=\left(\f{r^{1+(d-1)(\gamma-1)}}{1+(d-1)(\gamma-1)}{}_2F_1\left[a,b,c,\left(\f{r_{\star}}{r}\right)^{2(d-1)(\gamma-1)} \right]\right)^{r_{\star}}_{\epsilon},\quad {\rm for} ~\gamma\neq \f{d-2}{d-1}
    \ee
    where $a=\f{1}{2}, b=-\f{1+(d-1)(\gamma-1)}{2(d-1)(\gamma-1)},c=1-\f{1+(d-1)(\gamma-1)}{2(d-1)(\gamma-1)}$. The dependence on the $\epsilon$ goes as $\epsilon^{2-d+\gamma(d-1)}$.
    It is easy to notice that in the $\gamma\ra 0$ limit, it reduces to that written in eq(\ref{ee_hsv}) and  gives the precise entanglement entropy as obtained in \cite{Ryu:2006bv} and the $\epsilon^{2-d}$ behavior.  
    
    For $\gamma = \f{d-2}{d-1}$, it is easy to see from eq(\ref{s_ee_leading_order}) using eq(\ref{integrals}) that there arises a logarithmic dependence of the entanglement entropy as obtained in \cite{Huijse:2011ef}. For completeness, it comes out as
    \be
2 G_N    S_{EE}(\delta=1)=L^{d-2}\int^{r_{\star}}_{\epsilon} dr \f{r^{(\gamma-1)(d-2)+\gamma-1}}{\sqrt{1-\left(\f{r_{\star}}{r}\right)^{2(\gamma-1)(d-1)}}}\simeq L^{d-2}~Log\left(\f{2r_{\star}}{\epsilon} \right).
    \ee

   \paragraph{ Hypersurface at finite coupling but for $\lambda_2=\lambda_3=\lambda_4=0$:}
   
   Now, let us include the effect of the finite coupling, $\lambda_1$, for the diagonal  $d+1$ dimensional bulk spacetime, which is AdS. Upon doing a tedious but straight forward calculation, we find the following expressions
   \bea
   {\cal K}^1&=&\f{r\rq{}[(d-1)g_{xx}g\rq{}_{xx}+d r\rq{}^2 g_{rr}g\rq{}_{xx}-r\rq{}^2g_{xx}g\rq{}_{rr}-2r\rq{}\rq{}g_{xx}g_{rr}]}{2g_{xx}(g_{xx}+g_{rr}r\rq{}^2)^2}\nn
  {\cal K}^r&=&\f{2r\rq{}\rq{}g_{xx}g_{rr}-dr\rq{}^2g_{rr}g\rq{}_{xx}+r\rq{}^2g_{xx}g\rq{}_{rr}-(d-1)g_{xx}g\rq{}_{xx}}{2g_{rr}(g_{xx}+g_{rr}r\rq{}^2)^2}\nn   
  R_{x_1x_1}&=&\f{(d-2)}{4g^2_{xx}(g_{xx}+g_{rr}r\rq{}^2)}\Bigg[2r\rq{}^2 g_{xx}g\rq{}^2_{xx}+r\rq{}^4g_{xx}g\rq{}_{xx}g\rq{}_{rr}-2r\rq{}\rq{}g^2_{xx}g\rq{}_{xx}-2r\rq{}^2g^2_{xx}g\rq{}\rq{}_{xx}-
 \nn&& 2r\rq{}^4g_{xx}g_{rr}g\rq{}\rq{}_{xx}+r\rq{}^4g_{rr} g\rq{}^2_{xx}\Bigg]\nn
 R_{x_ix_j}&=&\f{\delta_{x_ix_j}}{4g_{xx}(g_{xx}+g_{rr}r\rq{}^2)^2}\Bigg[g_{xx}\bigg(r\rq{}^4g\rq{}_{xx}g\rq{}_{rr}-(d-5)r\rq{}^2g\rq{}^2_{xx}-2r\rq{}\rq{}g_{xx}g\rq{}_{xx}-2r\rq{}^2g_{xx}g\rq{}\rq{}_{xx}-
 \nn&&2r\rq{}^4g_{rr}g\rq{}\rq{}_{xx}\bigg) -(d-4)r\rq{}^4g_{rr}g\rq{}^2_{xx}\Bigg]\nn
 R&=&\f{(d-2)}{4g^2_{xx}(g_{xx}+g_{rr}r\rq{}^2)^2}\Bigg[2r\rq{}^4g_{xx}g\rq{}_{xx}g\rq{}_{rr}-(d-7)r\rq{}^2g_{xx}g\rq{}^2_{xx}-
 4r\rq{}\rq{}g^2_{xx}g\rq{}_{xx}-4r\rq{}^2g^2_{xx}g\rq{}\rq{}_{xx}\nn&&-4r\rq{}^4g_{xx}g_{rr}g\rq{}\rq{}_{xx}-
 (d-5)r\rq{}^4g_{rr}g\rq{}^2_{xx} \Bigg],
   \eea
   where $g\rq{}_{ij}=\p_r g_{ij}$ and $r\rq{}=\f{dr}{dx_1}$.
    In which case, the  equation of motion becomes
   \bea\label{eom_lambda1}
 &&4g^2_{xx}(g_{rr}+g_{xx}x\rq{}^2_1)\left[-g_{xx}(\p_rg_{rr})x\rq{}_1+d g_{rr}(\p_rg_{xx})x\rq{}_1+(d-1)g_{xx}(\p_rg_{xx})x\rq{}^3_1+2g_{rr}g_{xx}x\rq{}\rq{}_1\right]+\nn
 &&\lambda_1(d-3)(d-2)(\p_r g_{xx})\Bigg[3g_{xx}(\p_rg_{rr})(\p_rg_{xx})x\rq{}_1-(d-4)g_{rr}(\p_rg_{xx})^2 x\rq{}_1-(d-7)g_{xx}(\p_r g_{xx})^2 x\rq{}^3_1\nn && -4 g_{rr} g_{xx}(\p^2_r g_{xx})x\rq{}_1-4g^2_{xx}(\p^2_rg_{xx})x\rq{}^3_1-2g_{rr}g_{xx}(\p_rg_{xx})x\rq{}\rq{}_1+
 4g^2_{xx}(\p_rg_{xx})x\rq{}^2_1x\rq{}\rq{}_1\Bigg]=0,
   \eea
   where we have included  the contribution only from the first two terms of eq(\ref{assumption_EE}), i.e., have set $\lambda_2=\lambda_3=\lambda_4=0$. It is easy to notice that for $d=2$  the contribution from the induced Ricci scalar vanishes identically. The above equation of motion can be re-written as
   \be\label{sol_strip_lambda1}
   \f{d}{dr}\left[g^{\f{d-4}{2}}_{xx} x\rq{}_1\left(\f{4g_{rr}g^2_{xx}+4g^3_{xx}x\rq{}^2_1-\lambda_1(d-2)(d-3) (\p_rg_{xx})^2}{4(g_{rr}+g_{xx}x\rq{}^2_1)^\f{3}{2}}\right) \right] =0.
   \ee
   
 This gives a cubic equation in $x\rq{}^2_1$ and all the solution of it are not real.  In fact, the real solution for $x\rq{}_1$ is a huge expression and finding the exact analytical solution of the hypersurface, $x_1(r)$, is a daunting task. However, the derivative of the function, $x_1(r)$, which is real   given in  Appendix B.

   On computation of the holographic entanglement entropy functional, we find the entanglement entropy takes the following form
   \bea\label{ee_strip_lambda1}
2 G_N   &S_{EE}&=L^{d-2}\int dr \f{g^{\f{d-6}{2}}_{xx}}{4\left[g_{rr}+g_{xx}x\rq{}^2_1\right]^{\f{3}{2}}}\Bigg[4g^2_{xx}(g_{rr}+g_{xx}x\rq{}^2_1)^2+\lambda_1(d-2) \bigg(2g_{xx}g\rq{}_{xx}g\rq{}_{rr}-\nn&&
   (d-7)x\rq{}^2_1g_{xx}g\rq{}^2_{xx}+4x\rq{}_1x\rq{}\rq{}_1g^2_{xx}g\rq{}_{xx}-4x\rq{}^2_1 g^2_{xx}g\rq{}\rq{}_{xx}-4 g_{xx}g_{rr}g\rq{}\rq{}_{xx}-(d-5)g_{rr}g\rq{}^2_{xx} \bigg)\Bigg]\nn
   \eea
 
   Substituting the solution of $x\rq{}_1(r)$ to quadratic order in $\lambda_1$ from Appendix B into the above expression of the entanglement entropy, gives
   \bea\label{ee_2nd_order_lambda1}
2 G_N &S_{EE}&=L^{d-2}\int dr \f{\sqrt{g_{rr}(r)g^{d-2}_{xx}(r)}}{\sqrt{1-\f{c^2}{g^{d-1}_{xx}(r)}}}-L^{d-2}\lambda_1\f{(d-2)}{4} \int dr \Bigg[2c^2 g^{2}_{xx}(r)g\rq{}_{rr}(r)g\rq{}_{xx}(r)-\nn&&2g^{d+1}_{xx}(r)g\rq{}_{rr}(r)g\rq{}_{xx}(r)+6c^2g_{xx}(r)g_{rr}(r)g\rq{}^2_{xx}(r)+(d-5)g_{rr}(r)g^{d}_{xx}(r)g\rq{}^2_{xx}(r)+\nn&&4g_{rr}(r)g^{d+1}_{xx}(r)g\rq{}\rq{}_{xx}(r)-4c^2g_{rr}(r) g^{2}_{xx}(r) g\rq{}\rq{}_{xx}(r) \Bigg]/\left[ g^{3/2}_{rr}(r)g^{\f{d+6}{2}}_{xx}(r)\sqrt{1-\f{c^2}{g^{d-1}_{xx}(r)}}\right]+c^2\nn&&
   \f{(d-2)^2(d-3)}{32}
   L^{d-2}\lambda^2_1\int dr g\rq{}^2_{xx}\Bigg[g_{rr}(r)g_{xx}(r)g\rq{}^2_{xx}(r)\left(2c^2(3d+13)-3(d+9)g^{d-1}_{xx}(r)\right)+\nn
   &12&g^2_{xx}\left(c^2 -g^{d-1}_{xx}(r)\right)\left(g\rq{}_{rr}(r)g\rq{}_{xx}(r)-2g_{rr}(r)g\rq{}\rq{}_{xx}(r)\right)\Bigg]/\left[g^{5/2}_{rr}(r)g^{\f{3d+8}{2}}_{xx}(r)\sqrt{1-\f{c^2}{g^{d-1}_{xx}(r)}} \right]\nn
    &+&{\cal  O}\left(\lambda_1\right)^3
   \eea

    \subsection{Entanglement entropy}


  We can get the exact expression of the entanglement entropy upon substituting the solution from Appendix B into eq(\ref{ee_strip_lambda1}) for the geometry eq(\ref{generic_hsv}). Since, the form of $x\rq{}_1(r)$ is messy, let us use the leading order (in $\lambda_1$) form of it and perform   the computation of the entanglement entropy. In which case
   \bea
2 G_N  &S_{EE}&=L^{d-2}\int^{r_{\star}}_{\epsilon}dr \f{r^{(d-1)\gamma-1-(d-2)\delta} }{\sqrt{1-\left(\f{r}{r_{\star}} \right)^{2(d-1)(\delta-\gamma)}}}-\lambda_1L^{d-2}(d-2)\int^{r_{\star}}_{\epsilon}\f{dr}{\sqrt{1-\left(\f{r}{r_{\star}} \right)^{2(d-1)(\delta-\gamma)}}}\nn
   &&\Bigg[ (\gamma-\delta)[\gamma(d-3)-(d-1)\delta] r^{-1+(d-3)\gamma-(d-2)\delta}+2(\gamma-\delta)(2\gamma-\delta)\f{r^{-1-\gamma(d+1)+d\delta}}{r^{2(d-1)(\delta-\gamma)}_{\star}}\Bigg]\nn&-&\lambda^2_1(d-2)^2(d-3)(\gamma-\delta)^3L^{d-2}\int^{r_{\star}}_{\epsilon}\f{dr}{\sqrt{1-\left(\f{r}{r_{\star}} \right)^{2(d-1)(\delta-\gamma)}}}\times\nn&&\Bigg[\f{3}{2}[\gamma(d+5)-\delta(d+1)]\f{r^{d\delta-1-\gamma(d+3)}}{r^{2(d-1)(\delta-\gamma)}_{\star}}- [\gamma(3d+7)-\delta(3d+1)]\f{r^{(3d-2)\delta-1-\gamma(3d+1)}}{r^{4(d-1)(\delta-\gamma)}_{\star}}\Bigg]
   \nn&+&\cdots,\nn
   \eea
where the ellipses stands for higher order terms in $\lambda_1$. 
  On performing the integrals results in 
  \bea
2G_N  &S_{EE}&=\f{L^{d-2}}{[\gamma(d-1)-\delta(d-2)]}\times\nn
  &&\left(r^{\gamma(d-1)-\delta(d-2)}{}_2F_1\left[\f{1}{2},\f{\gamma(d-1)-\delta(d-2)}{2(d-1)(\delta-\gamma)},\f{d\delta-\gamma(d-1)}{2(d-1)(\delta-\gamma)},\left(\f{r}{r_{\star}}\right)^{2(d-1)(\delta-\gamma)}\right]\right)^{r_{\star}}_{\epsilon}\nn
  &-& \lambda_1L^{d-2}\f{(d-2)(\gamma-\delta)}{[\gamma(d-3)-(d-2)\delta]}[\gamma(d-3)-(d-1)\delta]\times\nn
  &&\left(r^{(d-3)\gamma-\delta(d-2)}{}_2F_1\left[\f{1}{2},\f{\gamma(d-3)-(d-2)\delta}{2(d-1)(\delta-\gamma)},\f{d\delta-\gamma(d+1)}{2(d-1)(\delta-\gamma)},\left(\f{r}{r_{\star}}\right)^{2(d-1)(\delta-\gamma)}\right]\right)^{r_{\star}}_{\epsilon}\nn
  &+&\lambda_1L^{d-2}\f{2(d-2)(2\gamma-\delta)(\gamma-\delta)}{[\gamma(d+1)-d\delta]}r^{2(\gamma-\delta)(d-1)}_{\star}
  \Bigg(r^{d\delta-\gamma(d+1)}\nn&&{}_2F_1\left[\f{1}{2},\f{d\delta-\gamma(d+1)}{2(d-1)(\delta-\gamma)},1+\f{d\delta-\gamma(d+1)}{2(d-1)(\delta-\gamma)},\left(\f{r}{r_{\star}}\right)^{2(d-1)(\delta-\gamma)}\right]\Bigg)^{r_{\star}}_{\epsilon}+\lambda^2_1(d-2)^2\nn&&
 \f{3}{2} (d-3)L^{d-2}(\gamma-\delta)^3\f{[\gamma(d+5)-\delta(d+1)]}{\gamma(d+3)-d\delta}r^{2(\gamma-\delta)(d-1)}_{\star} \Bigg(r^{d\delta-\gamma(d+3)}\nn&&{}_2F_1\left[\f{1}{2},\f{d\delta-\gamma(d+3)}{2(d-1)(\delta-\gamma)},\f{(3d-2)\delta-\gamma(3d+1)}{2(d-1)(\delta-\gamma)},\left(\f{r}{r_{\star}}\right)^{2(d-1)(\delta-\gamma)}\right]\Bigg)^{r_{\star}}_{\epsilon}-\lambda^2_1(d-2)^2 \nn&&(d-3)L^{d-2}(\gamma-\delta)^3\left[\f{\gamma(3d+7)-\delta(3d+1)}{\gamma(3d+1)-(3d-2)\delta}\right]r^{4(\gamma-\delta)(d-1)}_{\star} \Bigg(r^{(3d-2)\delta-\gamma(3d+1)}\nn&&{}_2F_1\left[\f{1}{2},\f{(3d-2)\delta-\gamma(3d+1)}{2(d-1)(\delta-\gamma)},\f{(5d-4)\delta-\gamma(5d-1)}{2(d-1)(\delta-\gamma)},\left(\f{r}{r_{\star}}\right)^{2(d-1)(\delta-\gamma)}\right]\Bigg)^{r_{\star}}_{\epsilon}\nn&+&\cdots.
  \eea
     It is worth mentioning that till this order in $\lambda_1$ do not give any logarithmic violation of the entanglement entropy for any choice of $\gamma$ with $\delta$ either $0$ or $1$ except $\gamma=\f{d-2}{d-1}$ with $\delta=1$.
 
    To get a feel of the entanglement entropy for $AdS$ solution, which means setting $\delta=1$ and $\gamma=0$,  restoring the AdS radius
   $R_0$
     \bea
     2 G_N&  S_{EE}&=\f{L^{d-2}R^{d-1}_0}{\epsilon^{d-2}}\left(\f{1-(d-1)(d-2)(\lambda_1/R^2_0)}{(d-2)}\right)-\nn&&L^{d-2}R^{d-1}_0\f{\sqrt{\pi}\Gamma\left(\f{d}{2(d-1)}\right)}{\Gamma\left(\f{1}{2(d-1)}\right)}r^{2-d}_{\star}\left(\f{1+(d-2)(d-3)(\lambda_1/R^2_0)}{d-2}\right)
  +{\cal O}(\lambda_1)^2,
  \eea 
  where we have taken the $\epsilon\ra 0$ limit and kept both the divergent and finite terms.
     In the absence of the higher derivative correction, the UV divergence was found to be $\epsilon^{2-d}$ in $d+1$ dimensional bulk spacetime \cite{Ryu:2006bv}. And, upon inclusion of the next  higher derivative term to the holographic entanglement entropy functional  gives the same power of the  UV divergent,  $\epsilon^{2-d}$. This is  observed, however only,  for $d=4$ in  \cite{deBoer:2011wk}. 
    
 We can re-express the above expression of the entanglement entropy in terms of the size $\ell$ and $R$. In which case, it reads as
    \bea
     2 G_N ~S_{EE}&=&\f{L^{d-2}R^{d-1}}{4\epsilon^{d-2}}\left(\f{4-(d+1)(d-1)(d-2)a\lambda_a}{(d-2)}\right)-\nn&&2^{d-2}L^{d-2}R^{d-1}\pi^{\f{d-1}{2}}\left(\f{\Gamma\left(\f{d}{2(d-1)}\right)}{\Gamma\left(\f{1}{2(d-1)}\right)}\right)^{d-1}\ell^{2-d}\left(\f{1+\f{3}{4}(d-1)(d-2)(d-3)a\lambda_a}{d-2}\right)
  \nn&&+{\cal O}(\lambda_a)^2.
  \eea  
     
      Note that $R_0$ and $R$ are the radii of AdS spacetime with and without the higher derivative correction, whose precise relation is given in section 6 and $\lambda_1=a\lambda_a R^2$.
        Upon comparing eq(\ref{assumption_EE}) with eq(C.24) of  \cite{Hung:2011xb}, we find $a=\f{2}{(d-2)(d-3)}$, after identifying the couplings $\lambda_a=\lambda_{there}$.
        
 It is easy to see that up to linear in the coupling, $\lambda_a$, the  the finite part and the singular part of the entanglement entropy obeys following differential equations
 \be
 \f{d}{d\ell}\left(\ell^{d-2}S^{fp}_{EE} \right)=0,\quad \f{d^{d-1}}{d\ell^{d-1}}\left(\ell^{d-2}S^{sp}_{EE} \right)=0,
 \ee
where  $S^{fp}_{EE}$ and $S^{sp}_{EE}$  stands for the finite part and singular part of the entanglement entropy, respectively. However, there exists another differential equation for the full $S_{EE}$
 \be
\ell\f{\p}{\p\ell}\left(\ell \f{\p}{\p\ell}+(d-2) \right)S_{EE}=0.
\ee
 
 Let us re-write the expression of the entanglement entropy as
 \bea
  2 G_N ~S_{EE}&=&\f{L^{d-2}R^{d-1}}{\epsilon^{d-2}}\left(\f{2d-6-(d+1)(d-1)\lambda_a}{2(d-2)(d-3)}\right)-\nn&&2^{d-2}L^{d-2}R^{d-1}\pi^{\f{d-1}{2}}\left(\f{\Gamma\left(\f{d}{2(d-1)}\right)}{\Gamma\left(\f{1}{2(d-1)}\right)}\right)^{d-1}\ell^{2-d}\left(\f{1+\f{3}{2}(d-1)\lambda_a}{d-2}\right)+{\cal O}(\lambda_a)^2
 \eea
   
 We can check that   the quantities defined in eq(\ref{plausible_flow_any_d}) holds good for $d=4$ when we take the coupling
to stay within the following window  $-\f{7}{36}\leq \lambda_a \leq \f{9}{100}$. This bound follows from the study of  causality and the positivity of the energy flux in \cite{Hofman:2008ar, Brigante:2008gz, Buchel:2009tt}.
     
       \subsection{Gauss-Bonnet combination}
         
 In this section, we shall find the form of the extremal hypersurface as well as  the entanglement entropy to leading order in the coupling $\lambda_i$ for a very special combination of the $\lambda_2,~\lambda_3$ and $\lambda_4$. In which case the holographic entanglement entropy functional takes the following form as in \cite{Hung:2011xb}
\be
4 G_N   S_{EE}=\int d^{d-1}\sigma~\sqrt{det( g_{ab})}\left[1+\lambda_1 R(g)+\Lambda\left( R^2(g)-4 R_{ab}(g)R^{ab}(g)+ R_{abcd}(g)R^{abcd}(g)\right)\right].
 \ee           
                   
  In what follows, we shall be interested to calculate the entanglement entropy when the entangling region is of the strip type. In which case, the induced geometry is as written in eq(\ref{embeddings_strip}). For  simplicity of   doing the computation, we shall fix the dimensionality of the bulk  spacetime.                     
                                  
  \paragraph{For  $ d=5$: } The bulk spacetime is a $5+1$ dimensional system whereas the induced metric is a $4$ dimensional spatial metric. 
  In this case, the Gauss-Bonnet combination, $ R^2(g)-4 R_{ab}(g)R^{ab}(g)+ R_{abcd}(g)R^{abcd}(g)$, is non-zero but topological. It gives a non-zero contribution to the action but  not to the equation of motion of the embedding field. This also agrees with the computation of  the equation of motion  of the embedding field, $X^M$, following from eq(\ref{GB_eom}). In this case, the equation of motion can also be obtained from eq(\ref{eom_lambda1}) by considering  $d=5$. Hence,
   it is easy to conclude that  the form of the hypersurface is same  as in the previous case. Moreover, the holographic entanglement entropy  becomes
  \bea
2 G_N  S_{EE}&=&L^3\int dr ~g^{3/2}_{xx}\sqrt{g_{rr}+g_{xx}x\rq{}^2_1}\Bigg[ 1+3\lambda_1\times\nn&&\left(\f{g\rq{}_{rr}g\rq{}_{xx}+g\rq{}^2_{xx}x\rq{}^2_1-2(g_{rr}+g_{xx}x\rq{}^2_1)g\rq{}\rq{}_{xx}+
  2g_{xx}g\rq{}_{xx}x\rq{}_1x\rq{}\rq{}_1}{2g_{xx}(g_{rr}+g_{xx}x\rq{}^2_1)^2}\right)+3\Lambda g\rq{}^2_{xx}\times \nn&&
  \left(\f{g_{rr}g\rq{}^2_{xx}+g_{xx}(g\rq{}_{rr}g\rq{}_{xx}+2g\rq{}^2_{xx}x\rq{}^2_1-2g_{rr}g\rq{}\rq{}_{xx})+2g^2_{xx}x\rq{}_1(g\rq{}_{xx}x\rq{}\rq{}_1-x\rq{}_1g\rq{}\rq{}_{xx})}{2g^4_{xx}(g_{rr}+g_{xx}x\rq{}^2_1)^3}\right)\Bigg]\nn
  \eea
 
 Substituting the solution, $x\rq{}_1$,  from Appendix B, for the AdS geometry and doing the $r$ integral resulting in the entanglement entropy to linear order in $\lambda_1$ and $\Lambda$
 \be
2 G_N S_{EE}=L^3R^4_0\left(\f{1-\f{12\lambda_1}{R^2_0}+\f{24\Lambda}{R^4_0}}{3\epsilon^3}\right)-11R^4_0\sqrt{\pi }L^3\f{(1+\f{6\lambda_1}{R^2_0})}{64 r^3_{\star}}\f{\Gamma\left(\f{-11}{8}\right)}{\Gamma\left(\f{1}{8}\right)},
 \ee
  where $\epsilon$ and $r_{\star}$ are the UV regulator and the point of maximum extension along the $r$ direction, respectively. It is interesting to observe that to the linear order in $\Lambda$, the Gauss-Bonnet coefficient  does not enter in the finite term of $S_{EE}$  whereas it enters in the divergent piece.
  
  \paragraph{For  $ d=6$: }  
    For this case, with the induced metric as written in eq(\ref{embeddings_strip}) gives the following  holographic entanglement entropy functional
    \bea
2 G_N   & S_{EE}&=L^4\int dr g^2_{xx}\sqrt{g_{rr}+g_{xx}x\rq{}^2_1}\Bigg[1+\lambda_1\times\nn&&\left(\f{-g_{rr}g\rq{}^2_{xx}+g_{xx}(x\rq{}^2_1g\rq{}^2_{xx}+2g\rq{}_{rr}g\rq{}_{xx}-4g_{rr}g\rq{}\rq{}_{xx})+4g^2_{xx}x\rq{}_1(
  g\rq{}_{xx}x\rq{}\rq{}_1-x\rq{}_1g\rq{}\rq{}_{xx}}{g^2_{xx}(g_{rr}+g_{xx}x\rq{}^2_1)^2}\right)-\nn&&3\Lambda g\rq{}^2_{xx}\times \nn&&
  \left(\f{3g_{rr}g\rq{}^2_{xx}+g_{xx}(4g\rq{}_{rr}g\rq{}_{xx}+7g\rq{}^2_{xx}x\rq{}^2_1-8g_{rr}g\rq{}\rq{}_{xx})+8g^2_{xx}x\rq{}_1(g\rq{}_{xx}x\rq{}\rq{}_1-x\rq{}_1g\rq{}\rq{}_{xx})}{2g^4_{xx}(g_{rr}+g_{xx}x\rq{}^2_1)^3}\right)\Bigg].\nn
    \eea
The equation of motion that follows takes the following form
\bea\label{eom_d_6}
&&2g^4_{xx}(g_{rr}+g_{xx}x\rq{}^2_1)^2(g_{xx}g\rq{}_{rr}x\rq{}_1-6g_{rr}g\rq{}_{xx}x\rq{}_1-5g_{xx}g\rq{}_{xx}x\rq{}^3-
2g_{rr}g_{xx}x\rq{}\rq{})+6\lambda_1 g^2_{xx}g\rq{}_{xx}\nn&&(g_{rr}+g_{xx}x\rq{}^2_1)(-3g_{xx}g\rq{}_{rr}g\rq{}_{xx}x\rq{}_1+2g_{rr}g\rq{}^2_{xx}x\rq{}_1-g_{xx}g\rq{}^2_{xx}x\rq{}^3+4g_{rr}g_{xx}g\rq{}\rq{}_{xx}x\rq{}_1+4g^2_{xx}g\rq{}\rq{}_{xx}x\rq{}^3+\nn&&
2g_{rr}g_{xx}g\rq{}_{xx}x\rq{}\rq{}-4g^2_{xx}g\rq{}_{xx}x\rq{}^2_1x\rq{}\rq{}_1)+3\Lambda g\rq{}^3_{xx}(5g_{xx}g\rq{}_{rr}g\rq{}_{xx}x\rq{}_1+2g_{rr}g\rq{}^2_{xx}x\rq{}_1+7g_{xx}g\rq{}^2_{xx}x\rq{}^3_1-\nn&&
8g_{rr}g_{xx}g\rq{}\rq{}_{xx}x\rq{}_1-8g^2_{xx}g\rq{}\rq{}_{xx}x\rq{}^3_1-2g_{rr}g_{xx}g\rq{}_{xx}x\rq{}\rq{}_1+
8g^2_{xx}g\rq{}_{xx}x\rq{}^2_1x\rq{}\rq{}_1)=0.
\eea
  
   This equation of motion can be re-written as 
   \be
   \f{d}{dr}\Bigg[\f{x\rq{}_1g^3_{xx}}{\sqrt{g_{rr}+g_{xx}x\rq{}^2_1}}-3\lambda_1\f{g_{xx}g\rq{}^2_{xx}x\rq{}_1}{(g_{rr}+g_{xx}x\rq{}^2_1)^{3/2}}+3\Lambda\f{g\rq{}^4_{xx}x\rq{}_1}{2g_{xx}(g_{rr}+g_{xx}x\rq{}^2_1)^{5/2}} \Bigg]=0.
 \ee    
  Upon solving the equation of motion, we find to linear order in $\lambda_1$ and $\Lambda$ as
  \be\label{velocity_gb_lambd1_Lambda}
  x\rq{}_1(r)=\f{c\sqrt{g_{rr}}}{\sqrt{g^6_{xx}-c^2g_{xx}}}+\f{3c\lambda_1 g\rq{}^2_{xx}}{g^2_{xx}\sqrt{g_{rr}}\sqrt{g^6_{xx}-c^2g_{xx}}}+3c\Lambda\f{(c^2-g^5_{xx})g\rq{}^4_{xx}}{2g^{3/2}_{rr}g^{9}_{xx}\sqrt{g^6_{xx}-c^2g_{xx}}},
\ee
where $c$ is a constant of integration and  is determined by demanding that as $r\ra r_{\star}$, $x\rq{}_1(r_{\star})$ diverges. Substituting this form of the solution into the action, results in
\bea
2 G_N&S_{EE}&=L^{4}\int dr \Bigg[\f{\sqrt{g_{rr}}g^{9/2}_{xx}}{\sqrt{g^5_{xx}-c^2}}-\nn&&\lambda_1\f{6c^2g_{rr}g\rq{}^2_{xx}+g_{rr}g^5_{xx}g\rq{}^2_{xx}+2g_{xx}(c^2-g^5_{xx})(g\rq{}_{rr}g\rq{}_{xx}-2g_{rr}g\rq{}\rq{}_{xx})}{g^{3/2}_{rr}g^{5/2}_{xx}\sqrt{g^5_{xx}-c^2}}\nn&-&
\Lambda \f{3g\rq{}^2_{xx}\sqrt{g^5_{xx}-c^2}}{2g^{5/2}_{rr}g^{19/2}_{xx}}(g_{rr}g\rq{}^2_{xx}(3g^5_{xx}-22c^2)-4g_{xx}(c^2-g^5_{xx})(g\rq{}_{rr}g\rq{}_{xx}-2g_{rr}g\rq{}\rq{}_{xx}))  \Bigg]\nn
\eea
 
Let us use the geometry of  AdS spacetime and do the $r$ integral from $\epsilon$, the UV cutoff, to the maximum extension in 
IR, $r_{\star}$. This gives 
\be
2 G_NS_{EE}=\f{L^4R^5_0}{4\epsilon^4}\left(1-\f{20\lambda_1}{R^2_0}+\f{120\Lambda}{R^4_0}\right)-\f{L^4R^5_0}{44 r^4_{\star}}\sqrt{\pi}\left(11+\f{132\lambda_1}{R^2_0}-\f{120\Lambda}{R^4_0}\right) \f{\Gamma\left(\f{3}{5} \right)}{\Gamma\left(\f{1}{10} \right)},
\ee
 where we have set the constant $c=R^5_0/r^5_{\star}$. In our previous studies, we found that the divergent term to the entanglement entropy goes as $\epsilon^{2-d}$ and the finite term has the following dependence, $r^{2-d}_{\star}$. For the Gauss-Bonnet term in the holographic entanglement entropy functional, we found this behavior again.
 
 Now computing the size $\ell$ from eq(\ref{velocity_gb_lambd1_Lambda}) to linear order in the couplings 
 \be
 \f{\ell}{2}=\sqrt{\pi}\left(11+\f{132\lambda_1}{R^2_0}-\f{120\Lambda}{R^4_0}\right)\f{\Gamma\left(\f{8}{5} \right)}{66\Gamma\left(\f{11}{10} \right)}~r_{\star}+\cdots,
 \ee
 where the ellipses stands for the terms higher order in the couplings. Re-expressing the entanglement entropy in terms of the size $\ell$
 \be
2 G_NS_{EE}=\f{L^4R^5_0}{4\epsilon^4}\left(1-\f{20\lambda_1}{R^2_0}+\f{120\Lambda}{R^4_0}\right)-\f{4L^4R^5_0}{11\ell^4}\pi^{5/2}\left(11+\f{660\lambda_1}{R^2_0}-\f{600\Lambda}{R^4_0}\right) \left(\f{\Gamma\left(\f{3}{5} \right)}{\Gamma\left(\f{1}{10} \right)}\right)^5+\cdots.
 \ee
Using the following relation  $R_0=R\left(1-3a \lambda_a-12X_2\Lambda_a \right)$, we can re-write
 \bea
2 G_NS_{EE}&=&\f{L^4R^5}{4\epsilon^4}\left(1-35a\lambda_a+300X_2\Lambda_a\right)-\nn&&\f{4L^4R^5}{11\ell^4}\pi^{5/2}\left(11+495 a\lambda_a-2460X_2\Lambda_a\right) \left(\f{\Gamma\left(\f{3}{5} \right)}{\Gamma\left(\f{1}{10} \right)}\right)^5+\cdots,
 \eea
 where we have set $\lambda_1=a\lambda_a R^2$ and $\Lambda=3X_2R^4\Lambda_a$. Upon comparing  eq(\ref{assumption_EE}) with eq(C.24) of  \cite{Hung:2011xb}, we find $a=1/6$ and $X_2=-1/24$  for $d=6$, after identifying the coupling $\lambda_a=\lambda_{there}$ and $\Lambda=\mu_{there}$.
From this expression of the entanglement entropy, it is easy to notice that $S_{EE}$ obeys the following differential equation for $d=6$

\be
\ell\f{\p}{\p\ell}\left(\ell \f{\p}{\p\ell}+(d-2) \right)S_{EE}=0 .
\ee

\section{Higher derivative  corrected extremal surfaces can\rq{}t penetrate the horizon }   
       
    The higher derivative  corrected equation of motion with $\lambda_2=\lambda_3=\lambda_4=0$ in any arbitrary dimension  can be re-written as
    \bea\label{extremal_surface_ee}
  &&4g^2_{xx}(g_{xx}+g_{rr}r\rq{}^2)\left[-(d-1)g_{xx}(\p_rg_{xx})+g_{xx}(\p_rg_{rr}) r\rq{}^2-d g_{rr}(\p_rg_{xx})r\rq{}^2+2g_{rr}g_{xx}r\rq{}\rq{}\right]+\nn   
   &&\lambda_1(d-3)(d-2)(\p_r g_{xx})\Bigg[-3g_{xx}(\p_rg_{rr})(\p_rg_{xx})r\rq{}^4+(d-4)g_{rr}(\p_rg_{xx})^2 r\rq{}^4+\nn&&(d-7)g_{xx}(\p_r g_{xx})^2 r\rq{}^2 +4 g_{rr} g_{xx}(\p^2_r g_{xx})r\rq{}^4+4g^2_{xx}(\p^2_rg_{xx})r\rq{}^2-2g_{rr}g_{xx}(\p_rg_{xx})r\rq{}^2r\rq{}\rq{}+\nn&&
 4g^2_{xx}(\p_rg_{xx})r\rq{}\rq{}\Bigg]=0,
    \eea
       
 Now we shall set up our logic following      \cite{Hubeny:2012ry} and show that after the inclusion of the higher derivative terms, i.e., with $\lambda_1\neq 0$, the extremal surfaces can\rq{}t penetrate the horizon, $r_h$.  The argument goes as follows. Eq(\ref{extremal_surface_ee}) essentially describes the equation of the hypersurface that ends on the boundary and goes deep into the bulk. Let us assume that it goes till $r=r_{\star}$, which is the deepest that it can go and then turns around and ends at the boundary. So at this point, $r_{\star}$, the derivative of the function $r$ with respect to $x_1$ vanishes, i.e., $\left( \f{dr}{dx_1}\right)_{r_{\star}}=0$. Putting this piece of information into eq(\ref{extremal_surface_ee}) gives
 \be\label{extremal_surface_cond}
 \left[g^2_{xx}\left(2g_{rr}r\rq{}\rq{}-(d-1) \p_r g_{xx}\right)+\lambda_1(d-2)(d-3)(\p_r g_{xx})^2r\rq{}\rq{}\right]_{r_{\star}}=0.
 \ee
 
 Before moving onto discuss the   $\lambda_1\neq 0$ case, let us first discuss  the $\lambda_1= 0$ case. Also, we want to make few assumptions on the metric components. Let there be a horizon at $r=r_h$,  if there exists more than one then  this  is  the outermost horizon. The quantity $g_{rr}$ changes sign as we go beyond $r_h$, i.e., $g_{rr} < 0$ for $r > r_h$\footnote{ Remember that the boundary is at $r=0$.} and assume  $\p_r g_{xx}$ is always negative\footnote{This is true for AdS spacetime.}. On summarizing the assumptions:

\[ g_{rr}(r) = \left\{
  \begin{array}{l l}
    +ve & \quad \textrm{for $r< r_h$ \quad (Outside the horizon)}\\
    -ve & \quad \textrm{for $r> r_h$\quad    (Inside the horizon)}
  \end{array} \right.\]
 
 and 
\be
g_{xx}(r) ~{\rm is~ +ve~ for~ any~r};\quad {\rm whereas  }~  g\rq{}_{xx}(r) ~{\rm is~ -ve~ for~ any~r}
 \ee 
 
 \paragraph{$\lambda_1= 0$: } Let us demand that the extremal surface goes deep into the bulk and has the maximum extension. It means we need to impose the above mentioned condition  along with  the further condition that the quantity $\left( \f{d^2r}{dx^2_1}\right)_{r_{\star}} <0$. For vanishing  $\lambda_1$, there follows from eq(\ref{extremal_surface_cond})  at $r=r_{\star}$ that
 \be
\left[ 2r\rq{}\rq{}g_{rr}-(d-1) g\rq{}_{xx}\right]_{ r_{\star}}=0.
 \ee

For  $r_{\star} < r_h$, i.e., the point $r_{\star}$ is outside the horizon. In which case, we can easily satisfy the above equation. To make things clear $\left( r\rq{}\rq{}g_{rr}\right)_{r_{\star}}$ is -ve whereas $g\rq{}_{xx}(r_{\star})$ is -ve.  

For  $r_{\star} > r_h$, i.e., the point $r_{\star}$ is inside the horizon. In which case, we can\rq{}t satisfy the above equation.
It means that the extremal surfaces cannot penetrate the horizon because the sum of two positive quantity can\rq{}t give zero. This is the argument put forward in  \cite{Hubeny:2012ry}.

  \paragraph{$\lambda_1\neq  0$: }     Once again imposing the condition that the    extremal surface goes deep into the bulk and has the maximum extension means we need to put the conditions as mentioned above along with      $\left( \f{d^2r}{dx^2_1}\right)_{r_{\star}} <0$.  The corresponding equation for the hypersurface at $r=r_{\star}$
  is given by
  \be
  \left[ g^2_{xx}(2r\rq{}\rq{}g_{rr}-(d-1) g\rq{}_{xx})+\lambda_1 (d-2)(d-3) r\rq{}\rq{}g\rq{}^2_{xx}\right]_{ r_{\star}}=0.
   \ee 
  
    As the hypersurface goes inside the horizon,      $r_{\star} >  r_h$,  we can satisfy the above equation   provided $\lambda_1 >0$. It means for positive values of $\lambda_1$, the spacelike hypersurfaces can cross the horizon. However, as we shall check explicitly for AdS black hole spacetime the condition  $\left( \f{d^2r}{dx^2_1}\right)_{r_{\star}} <0$ can\rq{}t be obeyed. Hence, the spacelike hypersurface can\rq{}t penetrate the horizon.
    
    Let us examine this in detail, at least to linear order  in $\lambda_1$, using the solution as written in Appendix B.   Recall that $\f{d^2r}{dx^2_1}=-\f{1}{x\rq{}^3_1}\f{d^2x_1}{dr^2}$. It means $\f{1}{x\rq{}^3_1}\f{d^2x_1}{dr^2}$ should be +ve at  $r=r_{\star}$ for positive coupling. Let us check this explicitly for the following geometry \cite{Cai:2001dz} with unit AdS radius
    \be\label{sol_cai}
    g_{xx}=\f{1}{r^2},~~~g_{rr}=\f{2\lambda_a}{r^2} \left[1-\sqrt{1-4\lambda_a(r/r_h)^d}\right]^{-1},~~~c=\f{1}{r^{d-1}_{\star}},~~~ \lambda_1=\f{2}{(d-1)(d-3)}\lambda_a
    \ee
    and it follows that 
    \be
   \left( \f{1}{x\rq{}^3_1}\f{d^2x_1}{dr^2}\right)_{r_{\star}}=\f{(d-1)}{r_{\star}}\left(1-\f{r^d_{\star}}{r^d_h}\right)-3\lambda_a\f{(d-1)}{r_{\star}}\left(1-\f{r^d_{\star}}{r^d_h}\right)^2.
    \ee
  
  Let us ask the question: Can there be an instance 
  for  $r_{\star} \geq r_h$ with positive dimension, $d$, as well as positive coupling, $\lambda_a >0$, for which  the quantity $  \left( \f{1}{x\rq{}^3}\f{d^2x_1}{dr^2}\right)_{r_{\star}}$ becomes +ve? The answer is none. This suggests  that the spacelike hypersurfaces can\rq{}t penetrate the horizon.

  Let us look at the penetration of the hypersurface from the point of view of the solution of the embedding field,  $x_1(r)$, from eq(\ref{sol_lambda1}) of Appendix B.  Using eq(\ref{sol_cai}), we find to leading order in $\lambda_a$
  \bea
  x_1(r)&=&c_1+\int \f{dr}{r^{d-1}_{\star}\sqrt{1-\left(\f{r}{r_{\star}}\right)^{2(d-1)}}}\f{r^{d-1}}{\sqrt{1-\left(\f{r}{r_h}\right)^{d}}}+\f{3\lambda_a}{2}\int dr \f{r^{1-d}_{\star}r^{d-1}}{\sqrt{1-\left(\f{r}{r_{\star}}\right)^{2(d-1)}}}\sqrt{1-\left(\f{r}{r_h}\right)^{d}}\nn
  &=& c_1+r_{\star}\int \f{dt}{\sqrt{1-t^{2(d-1)}}}\f{t^{d-1}}{\sqrt{1-\left(\f{r_{\star}}{r_h}\right)^d t^d}}+\f{3\lambda_a r_{\star}}{2}\int dx \f{t^{d-1}}{\sqrt{1-t^{2(d-1)}}}\sqrt{1-\left(\f{r_{\star}}{r_h}\right)^d t^d} \nn
   \eea
  where we have defined $t\equiv r/r_{\star}$ and $c_1$ is the constant of integration.  The constant of integration, $c_1$, is determined by imposing the boundary condition that $x_1(r_{\star})=0$. From the fig(1) it is easy to notice that $\f{x_1(r)-c_1}{r_{\star}}$ is positive, this means using the boundary condition at $r=r_{\star}$, we get $c_1$ to be a negative quantity. 
  
    One half of the figure of the full $U$ shaped  profile is plotted in fig(1).  Let us note that the hypersurface can go from $0$ to $r_{\star}$
   means $0\leq t \leq 1$. For $r_{\star} > r_h$ and $t$ approaching unity means $\sqrt{1-\left(\f{r_{\star}}{r_h}\right)^d t^d}$ becomes complex.
 This means inside the horizon there does not exist  any real valued solution, which suggests that the hypersurface can not penetrate the horizon. 
  
 \begin{figure}[t]\label{fig_profile}
   {\includegraphics[ width=10cm,height=6cm]{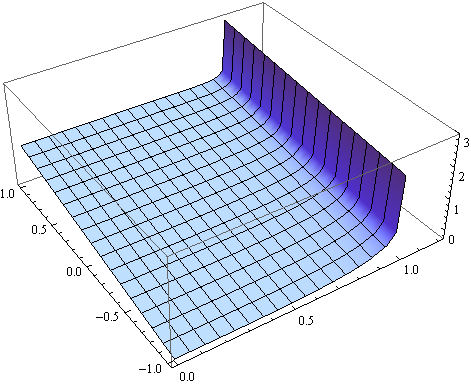} }
  \caption{
 The  graph is plotted for  $\f{x_1(r)-c_1}{r_{\star}}$ versus $r_{\star}/r_h$   in $d=4$.  The parameters  stay in the following range: $-1\leq \lambda_a \leq 1$ and $0\leq r_{\star}/r_h \leq 1.2$. }
\end{figure}

\section{Fluctuating geometry}

In this section, we would like to calculate the change in the entanglement entropy due to the background metric fluctuation, i.e., $G_{MN}\ra G_{MN}+\delta G_{MN}$ and keeping terms to linear order in $\delta G_{MN}$. Under such a change, the induced metric changes as $g_{ab}\ra g_{ab}+\delta g_{ab}$, where $\delta g_{ab}=\p_a X^M \p_b X^N \delta G_{MN}$. If we set $\lambda_{i+1}=0$ for $i\geq 1$, for simplicity, then the change in the entanglement entropy can be calculated from
\be
4G_N~\Delta S_{EE}=\int d^{d-1} \sigma \sqrt{det~g}\left[\f{1}{2}g^{cd}h_{cd}+\lambda_1\left( -R^{ab}h_{ab}+\nabla^a\nabla^b h_{ab}-g^{cd}\nabla^2 h_{cd}+\f{R}{2} g^{cd}h_{cd}\right) \right],
\ee
where $h_{ab}=\delta g_{ab}$. The indices are raised and lowered using $g_{ab}$ and its inverse. $\nabla_a$ is defined with respect to $g_{ab}$. In what follows, we are going to use a particular kind of metric fluctuation, namely that is used in the paper \cite{Bhattacharya:2012mi}, where the fluctuating metric is diagonal and asymptotes to the AdS geometry. Since, we did the computation of the entanglement entropy in full generality for the diagonal form of the metric as in (\ref{ee_2nd_order_lambda1}), which  suggests we can now do the computation for the fluctuating geometry as well.

\subsection{One parameter fluctuation with $\Lambda=0$}
For completeness, let us write down the complete  form of the metric with fluctuations and the bulk cosmological constant, $\Lambda_c$
\be\label{fluctuation_geometry_mass}
g_{tt}=\f{R^2_0}{r^2}(1-m r^d),\quad g_{xx}=\f{R^2_0}{r^2},\quad g_{rr}=\f{R^2_0}{r^2}(1+m r^d),\quad \Lambda_c=-\f{(d-1)(d-2)}{2R^2}
\ee
where we have restored AdS radius.  $m$ is a constant and  assumed to be very small, in the sense of \cite{Bhattacharya:2012mi}. In which case 
\bea
 \Delta S_{EE}&=&S_{EE}(m)-S_{EE}(m=0)=\f{m(d-1) L^{d-2}R^{d-1}_0}{32 G_N\sqrt{\pi}(d+1)}\ell^2 \f{\left(\Gamma\left( \f{1}{2(d-1)}\right)\right)^2}{\left(\Gamma\left( \f{d}{2(d-1)}\right)\right)^2} \f{\Gamma\left(\f{d}{(d-1)}\right)}{\Gamma\left( \f{d+1}{2(d-1)}\right)}\times\nn&&\left[1-3a(d-2)(d-3)\f{R^2}{ R^2_0}\lambda_a\right],
\eea
where we have set $\lambda_1=a R^2 \lambda_a$ with $a$  a constant.
Now, we shall move onto the computation of the energy (or mass) of the excited state using the AdS/CFT dictionary as worked out in  \cite{Balasubramanian:1999re}
\be
\Delta M=\int d^{d-1}x \sqrt{det(\sigma)_{ij}}~ N~u^M u^N T_{MN},
\ee
which in our case using $N=\sqrt{g_{tt}},~u^t=1/\sqrt{g_{tt}}$ and $ \sqrt{det(\sigma)_{ij}}=g^{\f{d-1}{2}}_{xx}=(R_0/r)^{d-1}$, gives
\be\label{exp_mass}
\Delta M=\int d^{d-1}x \left(\f{R_0}{r} \right)^{d-1}~\f{T_{tt}}{\sqrt{g_{tt}}}.
\ee

The $t-t$ component of the energy momentum tensor can be calculated from \cite{Dehghani:2006dh}
\bea
T_{tt}&=&\f{1}{8\pi G_N}\Bigg(K_{tt}+K g_{tt}+\f{d-1}{{\tilde R}}g_{tt}+2\lambda\Bigg[-\f{KK^2_{tt}}{g_{tt}}-\f{1}{3}\f{K^3_{tt}}{g^2_{tt}} +(d-1)\f{K_{tt}K^2_{xx}}{g^2_{xx}}-\nn&&K^2K_{tt}+(d-1)\f{Kg_{tt}K^2_{xx}}{g^2_{xx}}-\f{g_{tt}}{3}K^3-\f{2(d-1)}{3}\f{g_{tt}K^3_{xx}}{g^3_{xx}}\Bigg] \Bigg),
\eea
where $K_{tt}=-\f{ g\rq{}_{tt}}{2\sqrt{g_{rr}}},~K_{xx}=\f{ g\rq{}_{xx}}{2\sqrt{g_{rr}}}$ and $K=\f{ g\rq{}_{tt}}{2g_{tt}\sqrt{g_{rr}}}+\f{(d-1) g\rq{}_{xx}}{2g_{xx}\sqrt{g_{rr}}}$. The quantity ${\tilde R}$ is defined in such a way that as $\lambda\ra 0$, it approaches the size of the AdS spacetime\footnote{We use $R$ to denote the size of AdS spacetime without the higher derivative term, whereas $R_0$ with higher derivative term. The relation between them can be read out from \cite{Cai:2001dz}, $R^2_0=\f{R^2}{2}\left[1+\sqrt{1-\f{4(d-2)(d-3)\lambda}{R^2}}\right]$.} , i.e., $lim_{\lambda\ra 0}{\tilde R}\ra R$. It means we can write ${\tilde R}=R+\lambda R_1$ for small $\lambda$ and we shall
determine $R_1$ by demanding that $T_{tt}$ is not diverging as we approach the boundary, $r\ra 0$. The quantity $\lambda$ is same as $\alpha_2$ in the notation of  \cite{Dehghani:2006dh}.

Substituting all these ingredients into $T_{tt}$ gives
\be
T_{tt}=\f{(d-1)}{16 \pi G_N}m R_0 r^{d-2}-\f{(d-1)}{8\pi G_NR_0}m \lambda (d^2-5d+6)r^{d-2},
 \ee 
 where $R_1=\f{2(d^2-5d+6)}{3R_0}$. The mass comes as
 \be
 \Delta M=\f{(d-1)}{16\pi G_N}m \ell  L^{d-2}R^{d-1}_0-\f{(d-1)}{8\pi G_N}\lambda m \ell (d^2-5d+6)L^{d-2}R^{d-3}_0
 \ee
 
 On comparing with the bulk action and the holographic entanglement entropy functional of \cite{deBoer:2011wk} and \cite{Hung:2011xb} with ours, we find that
 $2\lambda= \lambda_1$. So the mass of the excited state becomes
 \be
 \Delta M=\f{(d-1)}{16\pi G_N}m \ell  L^{d-2}R^{d-1}_0-\f{(d-1)}{16\pi G_N}\lambda_1 m \ell (d^2-5d+6)L^{d-2}R^{d-3}_0
 \ee
 
 The AdS radius $R_0$ is related to $R$ as $R_0=R/\sqrt{f_{\infty}}$, where $f_{\infty}$ is found by solving $1-f_{\infty}+(a/2)(d-2)(d-3)\lambda_a f^2_{\infty}=0$. To leading order in $\lambda_a$, we take $R_0=R-(a/4)(d-2)(d-3) R \lambda_a$.
Finally, taking the ratio of the change in the entanglement with the mass gives
\bea\label{T_ent_lambda1}
\f{\Delta S_{EE}}{\Delta M}&=&\f{\sqrt{\pi}\left(\Gamma\left( \f{1}{2(d-1)}\right)\right)^2\Gamma\left( \f{1}{(d-1)}\right)}{2(d^2-1)\left(\Gamma\left( \f{d}{2(d-1)}\right)\right)^2\Gamma\left( \f{d+1}{2(d-1)}\right)}\ell\left[1-2 a \lambda_a (d^2-5 d+6)\right].
\eea

If we assume that there exists the following first law like of thermodynamics $T_{ent}~ \Delta S_{EE}=\Delta M$, then
\be
T_{ent}=c~ \ell^{-1},\quad {\rm with}\quad c=\f{2(d^2-1)\left(\Gamma\left( \f{d}{2(d-1)}\right)\right)^2\Gamma\left( \f{d+1}{2(d-1)}\right)}{\sqrt{\pi}\left(\Gamma\left( \f{1}{2(d-1)}\right)\right)^2\Gamma\left( \f{1}{(d-1)}\right)}\left[1+2a \lambda_a (d^2-5 d+6)\right].
\ee

The constant $a$ can be fixed by comparing eq(\ref{assumption_EE}) with  eq(C.24) of  \cite{Hung:2011xb} and there follows  $a=\f{2}{(d-2)(d-3)}$ for $\lambda_a=\lambda_{there}$. In which case 
\be
c=\f{2(d^2-1)\left(\Gamma\left( \f{d}{2(d-1)}\right)\right)^2\Gamma\left( \f{d+1}{2(d-1)}\right)}{\sqrt{\pi}\left(\Gamma\left( \f{1}{2(d-1)}\right)\right)^2\Gamma\left( \f{1}{(d-1)}\right)}\left[1+4 \lambda_a \right].
 \ee 
 At least for $d=4$, we find that the quantity $c$ is positive, which is expected. Recall, that the minimum value of $\lambda_a=-7/36$, follows from \cite{Hofman:2008ar, Brigante:2008gz, Buchel:2009tt}. In fact, for any $d$ with positive $T_{ent}$ requires us to set\footnote{However, for  $\lambda_a \leq -1/4$, we need to find an interpretation of $T_{ent}$.} $\lambda_a \geq -1/4$.
 
By doing the fluctuation to other parts of the metric component, the authors of \cite{Allahbakhshi:2013rda} found a modified first law like relation which involve both the \lq{}entanglement temperature\rq{} and \lq{}entanglement pressure\rq{}. For non-conformal theories the $T_{ent}$ is obtained in \cite{He:2013rsa} and \cite{Pang:2013lpa}.

\subsection{Fluctuation for non-zero $\lambda_1$ and $\Lambda$}

Let us calculate the change in the entanglement entropy and the mass due to the fluctuation in the geometry for $d=6$. Doing the one parameter  fluctuation, $m$, as done before in eq(\ref{fluctuation_geometry_mass}), we find that the change in entropy comes out as
\bea
\Delta S_{EE}&=&9L^4 R^5_0 m \ell^2 \left(187-\f{6732\lambda_1}{R^2_0}+\f{8040\Lambda}{R^4_0}\right)\left(\Gamma\left(\f{11}{10} \right) \right)^2 \times\nn&&
\left(\f{\Gamma\left(\f{1}{10} \right)\Gamma\left(\f{6}{5} \right)\Gamma\left(\f{8}{5} \right)\Gamma\left(\f{17}{10} \right)-\Gamma\left(\f{3}{5} \right)\Gamma\left(\f{7}{10} \right)\Gamma\left(\f{11}{10} \right)\Gamma\left(\f{11}{5} \right)}{1496 G_N \sqrt{\pi}\Gamma\left(\f{1}{10}\right)\Gamma\left(\f{7}{10}\right)\left( \Gamma\left(\f{8}{5}\right)\right)^3\Gamma\left(\f{17}{10} \right)}\right).
\eea

To this order, we can again read out the $t-t$ component of the energy-momentum tensor from \cite{Dehghani:2006dh}

\be
T_{ab}=\f{1}{8\pi G_N}\left(K_{ab}-KG_{ab} +2\lambda (3 J_{ab}-J G_{ab})+3{\tilde\Lambda}(5P_{ab}-PG_{ab})+\f{(d-1)}{{\tilde R}}G_{ab}\right).
\ee

We shall expand  ${\tilde R}=R+\lambda R_1+{\tilde\Lambda} R_2$ to linear order such that as we take the couplings to zero, we do get back the size of the AdS spacetime, $R$. The sizes $R_1$ and $R_2$ will be determined by demanding that $T_{tt}$ becomes finite as we approach the boundary. Or in the limit of  $m\ra 0$, the  $T_{tt}$ component should vanish as well \cite{Balasubramanian:1999re}. It gives $R_1=8/R_0$ and $R_2=-72/(5R^3_0)$.  Using all these ingredients into eq(\ref{exp_mass}), we find the mass becomes
\be
\Delta M=\f{5 L^4 R^5_0 m \ell }{16 \pi G_N}\left(1-\f{24\lambda}{R^2_0}+\f{72 {\tilde\Lambda}}{R^4_0} \right).
\ee

Now let us set the following relation between the bulk couplings $\lambda$ and ${\tilde \Lambda}$ with that appears on the holographic entanglement entropy functional $\lambda_1$ and $\Lambda$ following  \cite{Hung:2011xb}
\be
\f{\lambda}{b}=\f{\lambda_1}{a}\equiv\lambda_a R^2,\quad b=a/2,\quad \f{\Lambda}{X_1}=\f{{\tilde \Lambda}}{X_2}\equiv\Lambda_a R^4,\quad X_1=3X_2
\ee
with $a$ and $X_1$ are real numbers. The size of the AdS radii are related as
\be
R_0=\f{R}{\sqrt{f_{\infty}}},\quad {\rm where}\quad 1-f_{\infty}+\lambda_a f^2_{\infty}- \Lambda_af^3_{\infty}=0,
\ee
Note that while writing down such an equation, we have already used the relation between  $\lambda,~{\tilde \Lambda}$ and  $\lambda_1,~\Lambda$ as written above.
To linear order in the coupling we take $R_0=R\left(1-\f{1}{2} \lambda_a+\f{1}{2}\Lambda_a \right)$. Finally, taking the ratio 
\be
\f{\Delta S_{EE}}{\Delta M}=\left(\f{187-4488 a\lambda_a+3552X_1\Lambda_a}{13090}\right)\sqrt{\pi}~\ell~ \f{\Gamma\left(\f{1}{5}\right)\left( \Gamma\left(\f{1}{10}\right)\right)^2}{\Gamma\left(\f{7}{10}\right)\left( \Gamma\left(\f{3}{5}\right)\right)^2}.
\ee

The first law like of thermodynamics follows, $T_{ent}~ \Delta S_{EE}=\Delta M$, if we identify 
\be
T_{ent}=c~\ell^{-1},\quad {\rm with}\quad c=\f{70\Gamma\left(\f{7}{10}\right)\left( \Gamma\left(\f{3}{5}\right)\right)^2}{187\sqrt{\pi}\Gamma\left(\f{1}{5}\right)\left( \Gamma\left(\f{1}{10}\right)\right)^2} \left(187+4488 a\lambda_a-3552X_1\Lambda_a\right)
\ee

Let us fix the constants $a$ and $X_1$ by comparing eq(\ref{assumption_EE}) with  eq(C.24) of  \cite{Hung:2011xb}. It follows that $a=1/6$ and $X_1=-1/8$ for $d=6$, after identifying the couplings as $\lambda_a=\lambda_{there}$ and $\Lambda_a=\mu_{there}$. In which case
\be
 c=\f{70\Gamma\left(\f{7}{10}\right)\left( \Gamma\left(\f{3}{5}\right)\right)^2}{187\sqrt{\pi}\Gamma\left(\f{1}{5}\right)\left( \Gamma\left(\f{1}{10}\right)\right)^2} \left(187+748 \lambda_a+444\Lambda_a\right).
\ee

Positivity of $T_{ent}$ along with $\lambda_a \geq -1/4$ requires us to set $\Lambda\geq 0$.

\subsection{Two parameters fluctuation}

Now, we include the second parameter and study the change in the entanglement entropy as a function of these two parameters, $m$ and $q^2$, to  the  AdS geometry. Let us, write down the geometry with fluctuation as follows
\bea
g_{tt}&=&\f{R^2}{r^2}(1-m r^d+q^2 r^{2(d-1)}),\quad g_{xx}=\f{R^2}{r^2},\quad g_{rr}=\f{R^2}{r^2}(1+m r^d-q^2 r^{2(d-1)}),\nn  \Lambda_c&=&-\f{(d-1)(d-2)}{2R^2}.
\eea
The original  motivation to take such a form of the geometry is to compute the entanglement entropy with electric charges for RN-AdS
black hole. But, it is difficult, in practice, to  carry out the radial  integration involved, analytically, in the calculation of the entanglement entropy. Hence, we shall treat  $m$ and $q^2$ as small parameters 
\be
m~\ell^d\ll 1,\quad q^2 \ell^{2(d-1)}\ll 1.
\ee
With this kind of fluctuation, we shall compute the entanglement entropy. In fact, this computation is very easy to do, in the limit of vanishing of all the $\lambda_i$\rq{}s  in eq(\ref{ee_2nd_order_lambda1}).  The radial integral will be performed from the UV cutoff, $\epsilon$, to the turning point, $r_{\star}$. Moreover, the size $\ell$ is related to the turning point, $r_{\star}$.  We obtain the   entanglement entropy in terms of $\ell$ as

\bea
 S_{EE}&=&\f{L^{d-2}R^{d-1}}{2G_N(d-2)\epsilon^{d-2}}-L^{d-2}R^{d-1}\f{2^{d-3}\pi^{\f{d-1}{2}}}{(d-2)G_N}\ell^{2-d}\left(\f{ \Gamma\left(\f{d}{2(d-1)}\right)}{ \Gamma\left(\f{1}{2(d-1)}\right)}\right)^{d-1}\nn&&+\f{m(d-1) L^{d-2}R^{d-1}}{32 G_N\sqrt{\pi}(d+1)}\ell^2 \f{\left(\Gamma\left( \f{1}{2(d-1)}\right)\right)^2}{\left(\Gamma\left( \f{d}{2(d-1)}\right)\right)^2} \f{\Gamma\left(\f{d}{(d-1)}\right)}{\Gamma\left( \f{d+1}{2(d-1)}\right)}+q^2 L^{d-2} R^{d-1}\ell^d\f{\Gamma\left(\f{d}{2(d-1)}\right)}{8G_N \Gamma\left(\f{1}{2(d-1)}\right)}\nn&&
 \left(\f{1}{\sqrt{\pi}}\left(\f{\Gamma\left(\f{d}{2(d-1)}\right)}{ \Gamma\left(\f{1}{2(d-1)}\right)}\right)^{d-2} -(4\pi)^{\f{1-d}{2}}\right).
\eea

For $q=0$, it is easy to see that 
\be
\f{\ell^2\p^2_{\ell}S_{EE}-\ell \p_{\ell}S_{EE}}{\ell^2}=\ell \p_{\ell}\left(\ell^{-1}\p_{\ell}S_{EE} \right)<0,
\ee

for any $d\geq 3$. Whereas $\ell^2\p^2_{\ell}S_{EE}+\ell \p_{\ell}S_{EE}$ is not necessarily negative. So, also the quantity $\ell^2\p^2_{\ell}~S_{EE}=\ell \f{\p {\cal S}^{\Sigma}_3 }{\p\ell}$.

For $q\neq 0$, the following quantity 
\be
\f{\ell^2\p^2_{\ell}S_{EE}-\ell \p_{\ell}S_{EE}}{\ell^2}=\ell \p_{\ell}\left(\ell^{-1}\p_{\ell}S_{EE} \right)<0,
\ee

but only for $d\geq 5$.

\section{Conclusion and Open question}

The entanglement entropy is supposed to provide us the amount of classical/quantum information stored in a given region. The beautiful idea of \cite{Ryu:2006bv}  has led us    a new way to quantify it, using the celebrated AdS/CFT correspondence. In this paper, we used the Jacobson-Myers functional  \cite{Jacobson:1993xs} along with the prescription of \cite{Ryu:2006bv}   to compute the entanglement entropy of different kind of   systems. Such  systems are described by having different amount of symmetries and are called as Lifshitz solutions. 

As per \cite{Ryu:2006bv}, one of the important ingredient require to compute the entanglement entropy  is the hypersurface whose boundary coincides with the boundary of the given  region under study. The explicit  form of the hypersurface  is found using the prescription of \cite{Hubeny:2007xt} and because of its covariant nature, the hypersuface is independent of the  nature of the entangling region but  depends on the bulk couplings. The  form of the hypersurface is  obtained,  essentially, by  extremizing the Jacobson-Myers functional.
Apparently, it is not clear whether this form of the hypersurface  holds good even for time dependent geometries as well. 

Upon computing the  value of the action over the hypersurface which is considered to be of the shape of a strip gives us the desired result of the holographic entanglement entropy, $S_{EE}$. For a given size, $\ell$, the entanglement entropy obeys the following differential equation for AdS spacetime in $d\geq 3$
\be
\ell^2\f{\p^2 S_{EE}}{\p\ell^2}+(d-1)\ell \f{\p S_{EE}}{\p\ell}=0.
\ee

We have checked that even in the presence of the higher derivative terms to the holographic entanglement entropy functional, the entanglement entropy obeys the above mentioned,   simple looking, differential equation for the AdS spacetime with the following caveat. Since, the analytic computation of $S_{EE}$  to the higher orders in the couplings are very cumbersome. So, we have  computed $S_{EE}$ only to linear order in the couplings and checked the above mentioned differential equation. 

In \cite{Liu:2012eea}, a  useful quantity, \lq\lq{}renormalized entanglement entropy\rq\rq{} ${\cal S}^{\Sigma}_d$ is introduced, with which the authors have suggested to study the rate of  flow for a sphere type entangling region.  In our case, we have consider the scale $R$ in  \cite{Liu:2012eea} as $\ell$ and the surface, $\Sigma$, as the strip and  define to study the flow as $\ell \f{\p {\cal S}^{\Sigma}_d }{\p\ell} < 0$. We have checked    with the help of the differential equation obeyed by $S_{EE}$ and the exact form of $S_{EE}$ that such a quantity,  $\ell \f{\p {\cal S}^{\Sigma}_3 }{\p\ell} < 0$, holds true only for $d=3$. 

By going through an example of one parameter fluctuation, $m$, to the geometry, we have computed the entanglement entropy. From the result,   it is highly suggestive  to consider $\ell \p_{\ell}\left(\ell^{-1}\p_{\ell}S_{EE} \right)$ as the quantity that should give the rate of flow. As  the quantity $\ell \f{\p {\cal S}^{\Sigma}_3 }{\p\ell}$ is not necessarily negative.  However, by studying two parameter fluctuations, $m$ and $q$, we find that such a quantity becomes negative only for $d\geq 5$.

It is {\it a priori} not completely clear whether this gives the (complete) RG flow. Presumably, it is 
 interesting to include  the quantum corrections to the entanglement entropy of RT along the lines of \cite{Faulkner:2013ana, Barrella:2013wja}, and find the full RG flow structure, which we leave for future research.
 
We  have, also,  studied the first law like of thermodynamics for low excited states with higher derivative term. In which case,  the entanglement temperature $T_{ent}$  goes inversely with the size, $\ell$. In fact, the proportionality constant is a function of  the dimension, $d$, and the couplings.
\paragraph{Acknowledgment:} I am happy to thanks people who have helped me over the years. It is too difficult to write all the names. 
Thanks  to the computer center, IoP, Bhubaneswar for the help.

  \section{Appendix A: General Study of EE}
We shall show the finite term in the entanglement entropy of $d+1$ dimensional  hyperscale violating bulk spacetime
goes as $1/\ell^{d-2-\gamma(d-1)}$, where   $\ell$ is the size of the strip.
This we demonstrate by computing the entanglement entropy of the Lifshitz spacetime when the entangling region is of the strip type.    By doing a change of the coordinate, we re-express the Lifshitz geometry as a spacetime where the time coordinate scales linearly then we compute the entanglement entropy of this geometry. As expected, the entanglement entropy computed for these two cases  gives the same answer. 

To begin with, let us write down the geometry of the Lifshitz spacetime in $d+1$ dimensional spacetime with dynamical exponent $z$ as
\be
ds^2_{I,d+1}=R^2\left[ - \f{dt^2}{r^{2z}}+\f{dx^idx^j\delta_{ij}}{r^2}+\f{dr^2}{r^2}\right],
\ee
where the boundary is at $r=0$ and $R$ is the size of the bulk Lifshitz spacetime. It is easy to see that it has the following scaling symmetry
\be
r\ra \lambda~r,\quad t\ra \lambda^z~ t,\quad x^i\ra \lambda ~x^i.
\ee

We can re-write the geometry by doing the following change of coordinates $r^z=\rho,~t={\tilde t}/z,\quad x^i={\tilde x}^i/z,\quad R/z={\tilde R}$ as
\be
ds^2_{II,d+1}={\tilde R}^2\left[ - \f{d{\tilde t}^2}{\rho^{2}}+\f{d{\tilde x}^id{\tilde x}^j\delta_{ij}}{\rho^{2/z}}+\f{d\rho^2}{\rho^2}\right],
\ee
which has the following scaling symmetry
\be
\rho\ra \lambda^z \rho,\quad {\tilde t}\ra \lambda^z~ {\tilde t},\quad {\tilde x}^i\ra \lambda ~{\tilde x}^i.
\ee
The rationale behind taking such a non-linear scaling of the radial coordinate is that, we do not want to change the fact that we started out with  a spacetime which has a dynamical exponent $z$. 

Let us compute the entanglement entropy of these two spacetimes using the proposal of RT. According to the proposal the entanglement entropy is computed for a fixed time for which the $d-1$ dimensional spacelike hypersurface, $\gamma$,  extrmizes the area of this hypersurface.  Finally, the entanglement entropy is conjectured to take the following form, $S_{\gamma}=\f{Area(\gamma)}{4G_N}$.

Let us assume that the precise form of the hypersurface is determined by the function $r(x_1)$ and $\rho(x_1)$, in which case, the  induced metric for these two cases are
\bea
ds^2_{d-1}({\gamma_I})&=& R^2\left[ \left(\bigg(\f{dr}{dx_1}\bigg)^2+1\right)\f{dx^2_1}{r^2}+\f{dx^2_2+\cdots+dx^2_{d-1}}{r^2}\right],\nn
ds^2_{d-1}(\gamma_{II})&=&{\tilde R}^2\left[ \left(\bigg(\f{d\rho}{d{\tilde x}_1}\bigg)^2 \rho^{\f{2-2z}{z}}+1\right)\f{{\tilde x}^2_1}{\rho^{2/z}}+\f{d{\tilde x}^2_2+\cdots+d{\tilde x}^2_{d-1}}{\rho^{2/z}}\right].
 \eea 


Let us consider the entangling region is of the strip type. It means $0 \leq  x_1 \leq \ell$ and $-L/2\leq (x_2,\cdots,x_{d-1})\leq L/2$. In which case, the area becomes
\bea
{\cal A}({\gamma_I})&=& 2R^{d-1} L^{d-2}  \int \f{dr}{r^{d-1}} \sqrt{1+(dx_1/dr)^2} \nn
{\cal A}({\gamma_{II}})&=&2{\tilde R}^{d-1} {\tilde  L}^{d-2} \int \f{d\rho}{ \rho^{(d-1)/z}}\sqrt{\rho^{\f{2-2z}{z}}+(d{\tilde x}_1/d\rho)^2}.\quad {\tilde L}\equiv z L
\eea

Now we can extremize the area to find the hypersurface in both the cases, and are given as
\bea
\f{dx_1}{dr}=\f{(r/r_{\star})^{(d-1)}}{\sqrt{1-(r/r_{\star})^{2(d-1)}}},\quad
\f{d{\tilde x}_1}{d\rho}=\f{\rho^{\f{d-z}{z}} \rho_{\star}^{\f{-(d-1)}{z}}}{\sqrt{1-(\rho/\rho_{\star})^{\f{2(d-1)}{z}}}},
\eea
where $r_{\star}$ and $\rho_{\star}$ are determined as the place where the velocities diverges, i.e., $\left(\f{dx_1}{dr}\right)_{r_{\star}}\ra \infty$ and  $\left(\f{d{\tilde x}_1}{d\rho}\right)_{\rho_{\star}}\ra \infty$, respectively.  Substituting the solution into the area gives
\bea
{\cal A}({\gamma_I})&=&2R^{d-1} L^{d-2}  \int^{r_{\star}}_{\epsilon} \f{dr}{r^{d-1}} \f{1}{\sqrt{1-(r/r_{\star})^{2(d-1)}}},\nn
{\cal A}({\gamma_{II}})&=&2{\tilde R}^{d-1}  {\tilde L}^{d-2} \int^{\rho_{\star}}_{\varepsilon} \f{d\rho}{ \rho^{(d+z-2)/z}}\f{1}{\sqrt{1-(\rho/\rho_{\star})^{\f{2(d-1)}{z}}}},
\eea
where we have put a UV cutoff, $\epsilon$ and $\varepsilon$, to regulate the presence of divergences. As expected, for unit dynamical exponent there is no difference between these two areas. Let us use the following result for $r_{\star}>0$
\be
\int \f{dr}{r^n}\f{1}{\sqrt{1-(r/r_{\star})^{2m}}}=\f{r^{1-n}}{1-n}~{}_2F_1\left[ \f{1}{2},\f{1-n}{2m},1+\f{1-n}{2m},\left(\f{r}{r_{\star}}\right)^{2m}\right],\quad {\rm for}\quad  n\neq 1,
\ee
where ${}_2F_1[a,b,c,x]$ is the Hypergeometric function. 
It is very interesting to see that in both  cases for the area, the factor $\f{1-n}{m}$ are same. So, the final form of the area becomes
\bea\label{ee_hsv}
\f{{\cal A}({\gamma_I})}{2}&=&R^{d-1} L^{d-2} \left(\f{r^{2-d}}{2-d}~{}_2F_1\left[ \f{1}{2},\f{2-d}{2(d-1)},\f{d}{2(d-1)},\left(\f{r}{r_{\star}}\right)^{2(d-1)}\right] \right)^{r_{\star}}_{\epsilon},\nn
&=&\f{R^{d-1} L^{d-2}}{(d-2)}\f{1}{\epsilon^{d-2}}-\f{R^{d-1} L^{d-2}}{(d-2)} \sqrt{\pi}\f{\Gamma\left(\f{d}{2(d-1)}\right)}{\Gamma\left(\f{1}{2(d-1)} \right)}r^{2-d}_{\star}\nn
\f{{\cal A}({\gamma_{II}})}{2}&=&z{\tilde R}^{d-1} {\tilde L}^{d-2} \left(\f{\rho^{\f{2-d}{z}}}{2-d}~{}_2F_1\left[ \f{1}{2},\f{2-d}{2(d-1)},\f{d}{2(d-1)},\left(\f{\rho}{\rho_{\star}}\right)^{\f{2(d-1)}{z}}\right] \right)^{\rho_{\star}}_{\varepsilon},\nn
&=&\f{{\tilde R}^{d-1}{\tilde L}^{d-2}}{(d-2)}\f{z}{\varepsilon^{\f{d-2}{z}}}-z\f{ {\tilde R}^{d-1}{\tilde L}^{d-2}}{(d-2)} \sqrt{\pi}\f{\Gamma\left(\f{d}{2(d-1)}\right)}{\Gamma\left(\f{1}{2(d-1)} \right)}\rho^{\f{2-d}{z}}_{\star},
\eea
where we have used 
\be
{}_2F_1[a,b,c,1]=\f{\Gamma(c)\Gamma(c-a-b)}{\Gamma(c-a)\Gamma(c-b)},\quad {\rm for}\quad c\neq 0,-1,-2\cdots
\ee

Upon computing  $\ell$, the size of the strip along $x_1$, as a function of the turning point, $r_{\star}$ or $\rho_{\star}$, we find

\be
\ell/2=r_{\star} \sqrt{\pi}\f{\Gamma\left(\f{d}{2(d-1)}\right)}{\Gamma\left(\f{1}{2(d-1)}\right)},\quad \ell/2=\rho^{\f{1}{z}}_{\star} \sqrt{\pi}\f{\Gamma\left(\f{d}{2(d-1)}\right)}{\Gamma\left(\f{1}{2(d-1)}\right)}.
\ee

Now, substituting all these into the area
\bea\label{area_exp_l_ads}
\f{{\cal A}({\gamma_I})}{2}&=&\f{R^{d-1} L^{d-2}}{(d-2)}\f{1}{\epsilon^{d-2}}-\f{2^{d-2}R^{d-1} L^{d-2}}{(d-2)\ell^{d-2}} \pi^{\f{d-1}{2}}\left(\f{\Gamma\left(\f{d}{2(d-1)}\right)}{\Gamma\left(\f{1}{2(d-1)} \right)} \right)^{d-1},\nn
\f{{\cal A}({\gamma_{II}})}{2}&=&\f{ R^{d-1}L^{d-2}}{(d-2)}\f{1}{\varepsilon^{\f{d-2}{z}}}-z^{d-1}\f{2^{d-2}{\tilde R}^{d-1} L^{d-2}}{(d-2)\ell^{d-2}} \pi^{\f{d-1}{2}}\left(\f{\Gamma\left(\f{d}{2(d-1)}\right)}{\Gamma\left(\f{1}{2(d-1)} \right)} \right)^{d-1}.
\eea

Now it looks  that the finite piece  of the entanglement entropy are different for these two cases. As expected for $z=1$, they do give the same answer. Now the question is how to choose which one gives the correct entanglement entropy for the bulk Lifshitz spacetime? Actually, the finite pieces are not different. Recall, that in the second case, we have redefined the size of the Lifshitz spacetime and if we take care of that then they give precisely the same answer. Hence, there is no ambiguity.

Let us move to the computation of the entanglement entropy for the  hyperscaling violating theory. In particular, we are interested in the following bulk geometry
\be\label{geometry_hsv}
ds^2_{I,d+1}=R^2 r^{2\gamma}\left[ - \f{dt^2}{r^{2z}}+\f{dx^idx^j\delta_{ij}}{r^2}+\f{dr^2}{r^2}\right].
 \ee 
In order to have a boundary at $r=0$, we must take $\gamma <1$ with $z> 0$. This spacetime have the following scaling behavior
\be
r\ra \lambda~r,\quad t\ra \lambda^z~ t,\quad x^i\ra \lambda ~x^i,\quad ds\ra \lambda^{\gamma} ds.
\ee

  As considered previously, we can have another geometry with the same scaling behavior
\be
ds^2_{II,d+1}={\tilde R}^2\rho^{2\f{\gamma}{z}}\left[ - \f{d{\tilde t}^2}{\rho^{2}}+\f{d{\tilde x}^id{\tilde x}^j\delta_{ij}}{\rho^{2/z}}+\f{d\rho^2}{\rho^2}\right],
\ee
for which $\rho\ra \lambda^z \rho,\quad {\tilde t}\ra \lambda^z~ {\tilde t},\quad {\tilde x}^i\ra \lambda ~{\tilde x}^i,\quad ds\ra \lambda^{\gamma} ds.$
The induced geometries of the $d-1$ hypersurfaces becomes
\bea
ds^2_{d-1}({\gamma_I})&=& R^2 r^{2\gamma}\left[ \left(\bigg(\f{dr}{dx_1}\bigg)^2+1\right)\f{dx^2_1}{r^2}+\f{dx^2_2+\cdots+dx^2_{d-1}}{r^2}\right],\nn
ds^2_{d-1}(\gamma_{II})&=&{\tilde R}^2\rho^{2\f{\gamma}{z}}\left[ \left(\bigg(\f{d\rho}{d{\tilde x}_1}\bigg)^2 \rho^{\f{2-2z}{z}}+1\right)\f{d{\tilde x}^2_1}{\rho^{2/z}}+\f{d{\tilde x}^2_2+\cdots+d{\tilde x}^2_{d-1}}{\rho^{2/z}}\right].
\eea

Without giving the details, let us quote the area of the hypersurface 
\bea
\f{{\cal A}({\gamma_I})}{2}&=&R^{d-1} L^{d-2}  \int^{r_{\star}}_{\epsilon} \f{dr}{r^{(1-\gamma)(d-1)}} \f{1}{\sqrt{1-(r/r_{\star})^{2(1-\gamma)(d-1)}}},\nn
\f{{\cal A}({\gamma_{II}})}{2}&=&{\tilde R}^{d-1}  {\tilde L}^{d-2} \int^{\rho_{\star}}_{\varepsilon} \f{d\rho}{ \rho^{\f{d+z-2-\gamma(d-1)}{z}}}\f{1}{\sqrt{1-(\rho/\rho_{\star})^{\f{2(1-\gamma)(d-1)}{z}}}},
\eea

where the turning point, $r_{\star}$ and $\rho_{\star}$, are determined as the point where the velocity diverges, as found previously. The explicit form of the velocities are
\be
\f{dx_1}{dr}=\f{(r/r_{\star})^{(1-\gamma)(d-1)}}{\sqrt{1-(r/r_{\star})^{2(1-\gamma)(d-1)}}},\quad
\f{d{\tilde x}_1}{d\rho}=\f{\rho^{\f{d-z-\gamma(d-1)}{z}} \rho_{\star}^{\f{-(1-\gamma)(d-1)}{z}}}{\sqrt{1-(\rho/\rho_{\star})^{\f{2(1-\gamma)(d-1)}{z}}}}.
\ee

Doing the integrals we find 
\be
\ell/2=r_{\star} \sqrt{\pi}\f{\Gamma\left(\f{d-\gamma(d-1)}{2(1-\gamma)(d-1)}\right)}{\Gamma\left(\f{1}{2(1-\gamma)(d-1)}\right)},\quad \ell/2=\rho^{\f{1}{z}}_{\star} \sqrt{\pi}\f{\Gamma\left(\f{d-\gamma(d-1)}{2(1-\gamma)(d-1)}\right)}{\Gamma\left(\f{1}{2(1-\gamma)(d-1)}\right)},
\ee
for $\gamma\neq \f{d-2}{d-1}$ and $z\neq 1$.
Finally, the area integral becomes
\bea\label{s_ee_lifshitz_area_functional}
\f{{\cal A}({\gamma_I})}{2}&=&\f{R^{d-1} L^{d-2}}{(d-2-\gamma(d-1))}\f{1}{\epsilon^{d-2-\gamma(d-1)}}
-\f{R^{d-1} L^{d-2}}{(d-2-\gamma(d-1))(\ell/2)^{d-2-\gamma(d-1)}}\nn && \times\pi^{\f{(d-1)(1-\gamma)}{2}}\left(\f{\Gamma\left(\f{d-\gamma(d-1)}{2(1-\gamma)(d-1)}\right)}{\Gamma\left(\f{1}{2(1-\gamma)(d-1)} \right)} \right)^{(1-\gamma)(d-1)},\quad {\rm for}\quad \gamma\neq \f{d-2}{d-1}\nn
\f{{\cal A}({\gamma_{II}})}{2}&=&\f{ R^{d-1}L^{d-2}}{(d-2-\gamma(d-1))}\f{1}{\varepsilon^{\f{d-2-\gamma(d-1)}{z}}}
-\f{ R^{d-1} L^{d-2}}{(d-2-\gamma(d-1))(\ell/2)^{d-2-\gamma(d-1)}}\nn &&\times \pi^{\f{(d-1)(1-\gamma)}{2}}\left(\f{\Gamma\left(\f{d-\gamma(d-1)}{2(1-\gamma)(d-1)}\right)}{\Gamma\left(\f{1}{2(1-\gamma)(d-1)} \right)} \right)^{(1-\gamma)(d-1)}\quad {\rm for}\quad \gamma\neq \f{d-2}{d-1},\quad z\neq 1 .
\eea


Let us look at the $\gamma=\f{d-2}{d-1}$. This case has been analyzed earlier in  \cite{Huijse:2011ef}. Before doing the integral, let us look at the following integral
for $r_{\star}>0$
\bea\label{integrals}
\int \f{dr}{r}\f{1}{\sqrt{1-(r/r_{\star})^2}}&=& Log\left(\f{r}{1+\sqrt{1-(r/r_{\star})^2}} \right)\nn
\int \f{dr}{r^3}\f{1}{\sqrt{1-(r/r_{\star})^2}}&=&-\f{1}{2r^2}\sqrt{1-(r/r_{\star})^2}+\f{1}{2r^2_{\star}} Log\left(\f{r}{1+\sqrt{1-(r/r_{\star})^2}} \right),\nn 
\int \f{dr}{r^5}\f{1}{\sqrt{1-(r/r_{\star})^2}}&=&\f{3}{8r^4_{\star}} Log\left(\f{r}{1+\sqrt{1-(r/r_{\star})^2}} \right)-\sqrt{1-\f{r^2}{r^2_{\star}}}\left(\f{2r^2_{\star}+3r^2}{8r^4r^2_{\star}}\right).
\eea

Essentially, we are trying to find the cases, where there appears a log term in the area. For this choice of $\gamma=\f{d-2}{d-1}$, the area up to a divergent term   becomes
\be
{\cal A}(\gamma_I)=2R^{d-1}L^{d-2}  \int^{r_{\star}}_{\epsilon}\f{dr}{r}\f{1}{\sqrt{1-(r/r_{\star})^2}} \simeq 2R^{d-1}L^{d-2}  Log\left(2r_{\star} \right)+{\rm divergent\quad term},
\ee
which is the result found recently in \cite{Huijse:2011ef}.

  \section{Appendix B: The form of $x\rq{}_1(r)$ from eq(\ref{sol_strip_lambda1})}
We can re-write   eq(\ref{sol_strip_lambda1}) as $A_1(r) x\rq{}^6_1(r)+A_2(r) x\rq{}^4_1(r)+A_3(r) x\rq{}^2_1(r)+A_4(r)=0$, where
the $A_i$\rq{}s are  
  \bea
  A_1(r)&=&16g^3_{xx}(r)\left[g^{d-1}_{xx}(r)-c^2 \right],\quad A_4(r)=-16 c^2 g^3_{rr}(r)\nn
   A_2(r)&=&16g_{rr}(r)g^2_{xx}(r)\left[2 g^{d-1}_{xx}(r)-3c^2 \right]-8\lambda_1(d-2)(d-3) g^{d-1}_{xx}(r)g\rq{}^2_{xx}(r),\nn
   A_3(r)&=&16g^2_{rr}(r)g_{xx}(r)\left[ g^{d-1}_{xx}(r)-3c^2 \right]-8\lambda_1(d-2)(d-3) g_{rr}(r)g^{d-2}_{xx}(r)g\rq{}^2_{xx}(r)+\nn&&\lambda^2_1(d-2)^2(d-3)^2 g^{d-4}_{xx}(r)g\rq{}^4_{xx}(r)
  \eea
  
  and $c$ is a constant of integration. The real solution to $x\rq{}_1$ is
  \bea
  x\rq{}^2_1&=&-\f{A_2}{3A_1}+\nn
 &&  \f{2^{1/3}(A^2_2-3 A_1 A_3)}{3A_1\left(9A_1A_2A_3-2 A^3_2-27A^2_1A_4+\sqrt{(2A^3_2-9A_1A_2A_3+27A^2_1A^2_4)^2-4(A^2_2-3A_1A_3)^3}\right)^{1/3}}\nn
 &+&\f{\left(9A_1A_2A_3-2 A^3_2-27A^2_1A_4+\sqrt{(2A^3_2-9A_1A_2A_3+27A^2_1A^2_4)^2-4(A^2_2-3A_1A_3)^3}\right)^{1/3}}{3 \times 2^{1/3}A_1}.
  \eea
 We are not writing down the other two complex solutions and 
 to quadratic  order in $\lambda_1$, it  reads as
 \be\label{sol_lambda1}
 x\rq{}_1(r)=\f{c\sqrt{g_{rr}}}{\sqrt{g^d_{xx}-c^2 g_{xx}}}+\f{c(d-2)(d-3)\lambda_1g\rq{}^2_{xx}}{4g^2_{xx}\sqrt{g_{rr}}\sqrt{g^d_{xx}-c^2 g_{xx}}}-\f{c(d-2)^2(d-3)^2(3c^2g_{xx}-2g^d_{xx})g\rq{}^4_{xx}}{32g^{3/2}_{rr}g^{4+d}_{xx}\sqrt{g^d_{xx}-c^2 g_{xx}}} \lambda^{2}_1+{\cal O} \left(\lambda^{3}_1\right)
  \ee 
   
   For a geometry like eq(\ref{generic_hsv}), the solution to linear order in $\lambda_1$ reads as
   \bea
  \pm x_1(r)&=&c_1+\f{r^{d\delta-\gamma(d-1)}}{r^{(d-1)(\delta-\gamma)}_{\star}}\f{1}{[d\delta-\gamma(d-1)]}\times\nn&&{}_2F_1\left[ \f{1}{2},\f{d\delta-\gamma(d-1)}{2(d-1)(\delta-\gamma)},\f{\delta(3d-2)-\gamma(d-1)}{2(d-1)(\delta-\gamma)},\left(\f{r}{r_{\star}}\right)^{2(d-1)(\delta-\gamma)}\right]+\nn
  &&\lambda_1\f{(d-2)(d-3)(\gamma-\delta)^2}{[d\delta-\gamma(d-1)]}\f{r^{d\delta-\gamma(d-1)}}{r^{\delta(d-1)-\gamma(d+1)}_{\star}}\times\nn&&{}_2F_1\left[ \f{1}{2},\f{d\delta-\gamma(d+1)}{2(d-1)(\delta-\gamma)},\f{(3d-2)\delta-\gamma(3d-1)}{2(d-1)(\delta-\gamma)},\left(\f{r}{r_{\star}}\right)^{2(d-1)(\delta-\gamma)}\right]+{\cal O} \left(\lambda^{2}_1\right),\nn
   \eea
     where the constant $c_1$ is determined by imposing the condition $x_1(r=r_{\star})=0$.
     
     \section{Appendix C: The  couplings }
     
 In this section, we give the detailed relations of the couplings used in the paper.     We denote the bulk couplings as $(\lambda,~{\tilde \Lambda})$
 \be
 {\rm Bulk\quad  couplings:}\quad \lambda\equiv b\lambda_a R^2,\quad {\tilde \Lambda}\equiv X_2 \Lambda_a R^4.
 \ee

The couplings that appear in the holographic entangling entropy (HEE) functional are $( \lambda_1,~ \Lambda)$
\be
{\rm HEE\quad couplings:} \quad 
\lambda_1\equiv a \lambda_a R^2, \quad \Lambda\equiv X_1 \Lambda_a R^4.
\ee     
 where $R$ is the size of the AdS spacetime.  Now the relation between the bulk couplings and the HEE couplings are  \cite{Hung:2011xb}
 \be
 \lambda_1 =2\lambda,\quad \Lambda=3{\tilde \Lambda}\quad \Rightarrow\quad b=\f{a}{2},\quad X_1=3 X_2.
 \ee      
 
Also  it follows from  \cite{Hung:2011xb} that $a=\f{2}{(d-2)(d-3)}$ for any $d$ whereas $X_2=-1/24$ for $d=6$..
 
\end{document}